\documentclass[notitlepage,aps,twocolumn,superscriptaddress,longbibliography]{revtex4-1}
\usepackage{transparent}
\usepackage{graphicx}
\usepackage{xcolor}
\usepackage{amsmath}
\usepackage{amssymb}
\usepackage{todonotes}
\usepackage[unicode=true,colorlinks=true,citecolor=blue,urlcolor=blue]{hyperref}
\usepackage{float}
\usepackage{bm}
\usepackage{epsfig}
\usepackage{lipsum}
\usepackage{xcolor,soul}
\usepackage[normalem]{ulem}
\usepackage{physics}

\renewcommand {\phi}{{\varphi}}

\begin{document}

\title{Pole and zero edge state invariant for 1D non-Hermitian sublattice symmetry}
\author{Janet Zhong}
\affiliation{Department of Applied Physics, Stanford University, Stanford, California 94305, USA}
\author{Heming Wang}
\affiliation{Department of Electrical Engineering, Ginzton Laboratory, Stanford University, Stanford, California 94305, USA}
\author{Shanhui Fan}
\email{shanhui@stanford.edu}
\affiliation{Department of Applied Physics, Stanford University, Stanford, California 94305, USA}
\affiliation{Department of Electrical Engineering, Ginzton Laboratory, Stanford University, Stanford, California 94305, USA}

\begin{abstract}
There have been several criteria for the existence of topological edge states in 1D non-Hermitian two-band sublattice-symmetric tight-binding Hamiltonians. The generalized Brillouin zone (GBZ) approach uses the integration of the Berry connection over the GBZ contour in the complex wavevector space. An alternate `pole-zero' approach uses algebraic properties of the off-diagonal matrix elements of the sublattice-symmetric Hamiltonian in off-diagonal form. Both correctly predict the presence or absence of edge states, but there has not been an explicit proof of their equivalence. Here we provide such an explicit proof and moreover we extend the pole-zero approach so that it also applies for sublattice-symmetric models when the Hamiltonian is not in off-diagonal form. We give numerical examples for these invariants.
\end{abstract}
\date{\today}

\maketitle

\section{Introduction}
There has been substantial recent interest on the topology of band structures~\cite{hasan2019colloquium,chiu2016classification}. For Hermitian systems, the notion of bulk-edge correspondence states that the presence in a bulk system of a nonzero topological invariant, involving the integration of the Berry connection over the Brillouin zone, indicates the existences of topological edge states. This notion has been generalized to non-Hermitian systems~\cite{ashida2020nonhermitian,bergholtz2021exceptional,ding2022nonhermitian,lin2023topological}, where the integration needs to be performed over the generalized Brillouin zone (GBZ)~\cite{yao2018edge, yokomizo2019nonbloch}. In this paper, we refer to this approach as the GBZ approach.

As an alternative criterion for the existence of topological edge state, Lee and Thomale~\cite{lee2019anatomy} considered a specific class of one-dimensional two-band model having sub-lattice symmetry.  For this class of model, they consider the complex functions that describe the $z$-dependency of the off-diagonal matrix elements of the Bloch Hamiltonian, where $z = e^{ik}$ with $k$ being the wavevector. They show that the number of topological edge states is related to the poles and zeros of these complex functions. In this paper, we refer to this approach as the pole-zero approach. 

For non-Hermitian systems, the computation of the GBZ in general can be quite involved. Thus, the pole-zero approach of Ref.~\cite{lee2019anatomy} is interesting because no explicit computation of the GBZ is necessary. This is in contrast with several related works~\cite{yang2020nonhermitian,verma2023topological} which also relate the topological edge states to the behaviors of certain poles and zeros, but still require the knowledge of the GBZ in order to establish the criterion for the topological edge states. Both the GBZ approach and the pole-zero approach correctly describe the number of topological edge states and therefore must be equivalent to each other. However, there has not been a direct mathematical proof that establishes the equivalence of these two approaches. 

In this paper, we provide an explicit proof of the equivalence of the GBZ approach and the pole-zero approach for the description of the topological edge states. We use a Riemann-sphere interpretation of the eigenstates which provides some physical motivation for the pole-zero approach. We also generalize the pole-zero approach to a broader set of Hamiltonians with sublattice symmetry, such as SSH-Creutz models~\cite{zurita2021tunable,mccann2023catalog,lee2016anomalous}, by a modification of the formalism.  Our results clarifies certain theoretical aspects related to bulk-edge correspondence of edge states in non-Hermitian systems. Besides the GBZ and pole zero approach, some other approaches~\cite{bergholtz2021exceptional} use the delocalization transition of the biorthogonal polarization~\cite{kunst2018biorthogonal}, doubled Green's functions~\cite{borgnia2020nonhermitian} and real space wave-functions~\cite{song2019non}. Non-Hermitian topological edge states have since been experimentally demonstrated in photonics~\cite{weidemann2020topological,liu2024localization}, electric circuits~\cite{liu2021non-hermitian, helbig2020generalized,yuan2023nonhermitian}, metamaterials~\cite{fan2022hermitian,ghatak2020observation}, quantum optics~\cite{xiao2020nonhermitian} and more. Thus, as the 1D non-Hermitian sublattice symmetric model is arguably the simplest case exhibiting non-Hermitian topological zero-energy edge states, a deeper understanding of the topological invariants they correspond to may lead to new avenues in the interplay of edge states with other exotic phenomena~\cite{longhi2018nonhermitian,zhu2022anomalous,wan2023quantum,slim2024optomechanical,mcdonald2018phase,busnaina2023quantum,verma2024non,li2020critical,zhu2024versatile,liu2023topological,nehra2022topology,vyas2021topological,wu2020floquet,cao2021non,yang2024anatomy,zhu2024brief,edvardsson2019non,yao2018chern,hang2021nonlinear,tang2021topology,felbacq2023characterizing,hu2024non,yokomizo2024non} or more complex experimental systems.

The paper is organized as follows. In Sec.~\ref{sec:theory}, we summarize the main theoretical formulas and then introduce eigenvalue and eigenvector topology on a Riemann surface. We show the equivalence of the GBZ edge-state invariant and pole-zero invariant for one-dimensional, two-band, tight-binding models with sublattice symmetry where the Bloch matrix is in the off-diagonal form, i.e. the diagonal matrix elements are all zero. We then extend the formalism of the pole-zero approach so that the invariant applies for sublattice symmetric two-band models where the Bloch matrix is not in the off-diagonal form. In Sec.~\ref{sec:results}, we numerically demonstrate our pole-zero invariant and the GBZ invariant for a Hermitian SSH model, a generalized non-Hermitian SSH model with $t_3$ hoppings, a longer-range SSH with hopping range across three unit cells and a Hermitian and non-Hermitian SSH-Creutz model. We conclude in Sec.~\ref{sec:conclusion}.

\section{Theory}
\label{sec:theory}

\subsection{Theoretical setup and sublattice symmetry}
In this paper we consider two-band Hamltonians, which can be written in the wavevctor space as:
\begin{equation}
H(z)=\sum_{i=o,x,y,z}d_i(z) \sigma_i
\label{eq:H}
\end{equation}
 where $\sigma_i$ are Pauli matrices and the $d_i$'s are in general complex. The variable $z=e^{ik}$ is the phase factor with $k$ being the wavevector which is generally complex. Note that phase factor $z$ here differs from the subscript $z$ in $\sigma_z$ and $d_z$. The right eigenvalue equation for this Hamiltonian then reads:
 \begin{equation}
\left[\begin{array}{cc}d_o(z)+d_z(z) & d_x(z)-i d_y(z) \\ d_x(z)+i d_y(z) & d_o(z)-d_z(z)\end{array}\right]\left[\begin{array}{c}v_1^R \\ v_2^R\end{array}\right]=E(z)\left[\begin{array}{c}v_1^R \\ v_2^R\end{array}\right]
\label{Eq:characteristic}
\end{equation}
where $E(z)$ is the energy eigenvalue which has dispersion relation of 
\begin{equation}
    E_{\pm}(z) = d_o(z) \pm \sqrt{d_x^2(z)+d_y^2(z)+d_z^2(z)}
    \label{Eq:E}
\end{equation}
as can be obtained from the characteristic polynomial of $H(z)$, i.e $\det(H-E \mathbb{I})=0$.

We now consider a sublattice-symmetric Hamiltonian~\cite{yao2018edge,yokomizo2022nonhermitian,lee2019anatomy}. A Hamiltonian has a sublattice symmetry when it satisfies~\cite{wang2024nonhermitian,kawabata2019symmetry}
\begin{equation}
\Gamma H(z) \Gamma^{-1}=\Gamma H(z) \Gamma^{\dagger}=-H(z),
\label{Eq:sublattice}
\end{equation}
where $\Gamma$ is a unitary and Hermitian matrix~\cite{kawabata2019symmetry,asboth2015ashort}. Without loss of generality, we choose $\Gamma=\sigma_z$. From Eq.~\eqref{Eq:sublattice}, this implies that $d_{o}(z)=-d_{o}(z)$  and $d_{z}(z)=-d_{z}(z)$, which then implies $d_{o}(z)=0$ and $d_{z}(z)=0$. Thus, $H(z)$ can be written in the \textit{off-diagonal form}:
\begin{equation}
H(z)=\left[\begin{array}{cc}
0 & H_{ab}(z) \\
H_{ba}(z) & 0
\end{array}\right]
\label{eq:offdiagonal}
\end{equation}
The Hamiltonian of the form of Eq.~\eqref{eq:offdiagonal} is widely used for the study of bulk-edge correspondence in non-Hermitian~\cite{yao2018edge,lee2019anatomy,yokomizo2019nonbloch,kunst2018biorthogonal} and Hermitian~\cite{chiu2016classification} systems.

We note that the sublattice symmetry we consider in this paper differs from the non-Hermitian chiral symmetry as defined by~\cite{kawabata2019symmetry}:
\begin{equation}
\Gamma H^\dagger(z) \Gamma^{-1}=\Gamma H^\dagger(z) \Gamma^{\dagger}=-H(z).
\label{Eq:chiral}
\end{equation}
To contrast the two symmetries, again choose $\Gamma=\sigma_z$. Then from Eq.~\eqref{Eq:chiral}, we get the constraints $d_{o}(z)=-d_{o}^*(z)$ and $d_{z}(z)=-d_{z}^*(z)$ which implies $d_{o}(z)$ and $d_{z}(z)$ are imaginary but not necessarily zero. Thus, in non-Hermitian systems, chiral symmetry does not imply that the Hamiltonian takes the off-diagonal form. As a caveat, the notation about various symmetries has not been standarized in the literature. A number of papers refer to the sublattice symmetry as defined by Eq.~\eqref{Eq:sublattice} as chiral symmetry~\cite{jiang2018topological,liu2023topological,yao2018edge,kunst2018biorthogonal}. Here we follow the notation of Ref.~\cite{kawabata2019symmetry}.  

For subsequent use, we define the eigenvector ratio $M(z) = v_1^R(z)/v_2^R(z)$. From Eq.~\eqref{Eq:characteristic}, we have:
\begin{equation}
M(z)=\frac{d_x(z)-i d_y(z)}{E(z)-d_o(z)-d_z(z)}=\frac{E(z)-d_o(z)+d_z(z)}{d_x(z)+i d_y(z)}.
\label{Eq:Mdef}
\end{equation}
Using $M(z)$ allows us to study the eigenvector topology using a single complex number $M(z)$ rather than the two-component Bloch vector.

\subsection{Summary of main theoretical results}

When a real-space lattice as described by the Hamiltonian of Eq.~\eqref{eq:H} is truncated, it may support topological edge states. Our paper concerns the theoretical criterion for the existence of edge states as deduced
from topological properties derived from the bulk eigenstates. For non-Hermitian systems, the bulk eigenstates we use are the open-boundary solutions when the number of the unit cells $N \to \infty$. These eigenstates are charaterized by $z$-values that form the generalized Brillouin zone (GBZ)~\cite{yokomizo2019nonbloch,yao2018edge}. The GBZ can be obtained by analyzing the characteristic equation $\det(H(z)-E \mathbb{I} ) = 0$ and defines a contour on the complex $z$-plane, $\mathcal{C}_{\text{gbz}}$ (see Sec.~\ref{sec:proof} for more details). The GBZ invariant for the bulk-edge correspondence of a two-band Hamiltonian in the off-diagonal form is~\cite{yao2018edge,yokomizo2019nonbloch,verma2023topological}:

\begin{equation}
W_{\text{gbz}} \equiv \oint_{\mathcal{C}_{\text{gbz}}} \frac{1}{ 2 \pi} \frac{d}{d z} \arg (M^2(z)) d z
\label{Eq:gbz}
\end{equation}
where $W_{\text{gbz}}$ is a winding number. Though written in a different way by using $M(z)$, Eq.~\eqref{Eq:gbz} is equivalent to twice the non-Bloch winding number~\cite{yokomizo2019nonbloch,yao2018edge} (see Sec.~\ref{sec:proof}). It was shown in Ref.~\cite{yao2018edge,yokomizo2019nonbloch,yang2020nonhermitian,verma2023topological} that $|W_{\text{gbz}}|$ is equal to the number of topological edge states. 

Instead of Eq.~\eqref{Eq:gbz}, Ref.~\cite{lee2019anatomy} proposed an alternative way to obtain the number of topological edge states using algebraic properties involving the roots of the matrix elements of $H_{ab}(z)$ and $H_{ba}(z)$. We restate the theorem from Ref.~\cite{lee2019anatomy} in a different way, but we prove that our statement implies the statement from Ref.~\cite{lee2019anatomy} in Appendix.~\ref{sec:leepolezero}. Note that $H_{ab}(z)$ and $H_{ba}(z)$ are Laurent polynomials in $z$. Let $m>0$ be the highest negative power of $z$ in $H_{ab}(z)$, let $n>0$ be the highest negative power of $z$ in $H_{ba}(z)$ and let $\mu = \max\{m,n\}$. $\mu$ is then the range of unit cells (not sublattice sites) the longest hopping term reaches across. Then 
the number of edge states is equal to the absolute value of the following pole-zero invariant:
\begin{equation}
 W_{\text{pz}}=\# M^2_{\text{zeros} }- \# M^2_{\text{poles}}
\label{Eq:polezero}
\end{equation}
for the first 2$\mu$ zeros and poles of $M^2(z)$ where the zeros and poles are sorted in increasing order of  the $|z|$ magnitude at which they occur. Note that we count each zero and pole according to its multiplicity. 

The benefit of using poles and zeros of $M^2(z)$ in Eq.~\eqref{Eq:polezero} is that it can be easier to calculate than the GBZ. The main result of this paper that we will prove 
\begin{equation}
     W_{\text{pz}} =  W_{\text{gbz}}.
     \label{Eq:WgbzWpz}
\end{equation}

 We do not try to verify or prove Eq.~\eqref{Eq:polezero} from first principles as this is done in Ref.~\cite{lee2019anatomy}. Instead, the goal of this paper is to show the equivalence of Ref.~\cite{lee2019anatomy} to Ref.~\cite{yao2018edge,yokomizo2019nonbloch} (Sec.~\ref{sec:proof}). By bridging these two previously independently defined topological edge state invariants, we hope analytical insights from each invariant can now be connected, resulting in a deeper overall understanding of topological edge states in 1D non-Hermitian sublattice symmetry. Our work goes beyond Ref.~\cite{lee2019anatomy,yao2018edge,yokomizo2019nonbloch} by extending the pole-zero invariant and an interpretation of the GBZ invariant on the Riemann sphere for sublattice symmetric two-band models not in off-diagonal form (Sec.~\ref{sec:gen}).

\subsection{Proof of equivalence of GBZ and pole-zero invariant}
\label{sec:proof}
Consider an arbitrary one-dimensional tight-binding model with $d$ sublattice sites in a unit cell. The characteristic polynomial is:
\begin{align}
    \det(H(z)-E(z) \mathbb{I})=0.
    \label{Eq:char}
\end{align}
Eq.~\eqref{Eq:char} is a polynomial of degree $d$ in $E$ and a Laurent polynomial in $z$ with exponents $z^l$ where $-p \leq l \leq q$. Here $p$ and $q$ are given by the hopping range  to the left and right directions, and the choices of coupling parameters~\cite{zhang2020correspondence,wang2024nonhermitian}. By definition, $p$ is also the magnitude of the largest negative exponent of $z$ in the characteristics polynomial as defined above. There are $p+q$ solutions for $z$ at every $E$, and we number them in increasing order of magnitude $|z_1| \leq |z_2| \leq \hdots |z_{p+q}|$. Then the GBZ are the $z_p$ and $z_{p+1}$ solutions for Eq.~\eqref{Eq:char} when $|z_p| = |z_{p+1}|$. For more details and more general cases, see Ref.~\cite{wang2024nonhermitian}.

Now let us restrict to a two-band model so that $d=2$. For the rest of this section, we further assume that the Hamiltonian has sublattice symmetry and is already in the off-diagonal form. Then the characteristic polynomial in Eq.~\eqref{Eq:char} gives 
\begin{equation}
    E^2(z) = d_x^2(z) +d_y^2(z) .
    \label{Eq:twoband}
\end{equation}
Note that $d_x^2(z) +d_y^2(z) $ is again a Laurent polynomial in $z$. We also define a single-band Hamiltonian $Q(z)$ given by~\cite{longhi2019probing}: 
\begin{equation}
    Q(z) = d_x^2(z) +d_y^2(z) 
    \label{Eq:singleband}
\end{equation}
for which the GBZ is still given by $|z_p| = |z_{p+1}|$. Thus, the two-band model in Eq.~\eqref{Eq:twoband} is related to the single-band model via $E(z)=\pm \sqrt{Q(z)}$ and the GBZ of the two-band sublattice-symmetric model is two copies of the GBZ of the single-band model~\cite{wang2024nonhermitian}. Note that here we used $d_o(z)=0$ which arises from $H(z)$ being in the off-diagonal form as a result of sublattice symmetry. For general two-band models with $d_o(z) \neq 0$, one can no longer construct in a simple way a one-band model that has the same GBZ, as we have done here~\cite{longhi2019probing}.

Let us give a bit more background on where Eq.~\eqref{Eq:gbz} comes from. The non-Bloch winding number $W$~\cite{yao2018edge,yokomizo2019nonbloch} is given by the winding of the off-diagonal matrix elements:
\begin{equation}
    \begin{gathered}
W_1=\frac{1}{2 \pi} \oint_{\mathcal{C}_{\text{gbz}}} \arg H_{a b}(z) d z \\
W_2=\frac{1}{2 \pi} \oint_{\mathcal{C}_{\text{gbz}}} \arg H_{b a}(z) d z \\
W=W_1-W_2.
\label{Eq:Wwinding}
\end{gathered}
\end{equation}
where $H_{ab}(z) = d_x(z)-id_y(z), H_{ba}(z) = d_x(z)+id_y(z)$. Note that for a two-band model with sublattice symmetry, there are two copies of the GBZ laying on top of each other on the $z$-plane~\cite{wang2024nonhermitian}. Let us call each of the copies a subGBZ loop. Here, the integration contour $\mathcal{C}_{\text{gbz}}$ in Eq.~\eqref{Eq:gbz}, $W_1$ and in $W_2$ is an integral over just one of the two subGBZ loops, which is similar to taking the integral over the BZ of the Berry connection for a single band only for a two-band Hermitian Hamiltonian. Eq.~\eqref{Eq:gbz} is single-valued and analytic in $\arg(M^2(z))$ along the GBZ contour so long as the GBZ contour does not intersect zeros or poles of $M^2(z)$ on the $z$-plane. Thus, taking the integration along one of the two copies is well-defined regardless of the underlying GBZ energy band braiding topology~\cite{fu2024braiding,li2022topological,wang2024nonhermitian,hu2021knots,wojcik2020homotopy,li2021homotopical}. $|W|$ is equal to the total number of edge states~\cite{imura2019generalized,yokomizo2019nonbloch,yao2018edge,yang2020nonhermitian} (note that some references include a factor of 1/2 in Eq.~\eqref{Eq:Wwinding}~\cite{imura2019generalized, yokomizo2019nonbloch,yang2020nonhermitian}). Eq.~\eqref{Eq:gbz} is just the Hermitian $Q$-matrix invariant~\cite{chiu2016classification,asboth2015ashort} when the model is Hermitian. Eq.~\eqref{Eq:gbz} is not analytic if the GBZ contour intersects with zeros or poles of $M^2(z)$ on the $z$-plane. Such cases would correspond to gap closing points~\cite{fu2023anatomy} where the OBC bands touch at $E=0$ (see Appendix.~\ref{sec:branch}).

When $H(z)$ is in the off-diagonal form, we can write
\begin{equation}
    M^2(z)=\frac{H_{a b}(z)}{H_{b a}(z)}.
\end{equation}
Then since
\begin{equation}
    \arg \left(M^2(z)\right)=\arg \left(\frac{H_{a b}(z)}{H_{b a}(z)}\right)=\arg H_{a b}(z)-\arg H_{b a}(z).
\end{equation}
we have
\begin{equation}
    W= \frac{1}{2 \pi} \oint_{\mathcal{C}_{g b z}} \arg M^2(z) d z = W_{\text{gbz}}
        \label{Eq:Msquared}
\end{equation}
As $M^2(z)$ is the ratio of two Laurent polynomials, it is a meromorphic function. From the argument principle, Eq.~\eqref{Eq:Msquared} states that $W$ is equal to the number of zeros minus the number of poles of $M^2(z)$ within the GBZ on the $z$ plane. While this statement already resembles Eq. \eqref{Eq:polezero}, the important difference here is that $W_{\text{gbz}}$ counts the poles and zeros within the GBZ, while $W_{\text{pz}}$ counts the poles and zeros within a circle centered on $z=0$ that encloses the first $2\mu$ points.

To prove Eq.~\eqref{Eq:WgbzWpz}, we 
note that the single-band Hamiltonian $Q(z)$ can be written as 
\begin{equation}
    Q(z) = \frac{P_{p+q}(z)}{z^p}
\end{equation}
where $P_{p+q}(z)$ is a polynomial in $z$ of degree $p+q$. Refs.~\cite{zhang2020correspondence,okuma2020topological} states that the GBZ encloses the first $p$ zeros of $P_{p+q}(z)$. The proof is quite extensive and can be found in the Supplementary of Ref.~\cite{zhang2020correspondence}.

The zeros of $P_{p+q}(z)$ are also either poles or zeros of $M^2(z)$. This is a direct consequence of
\begin{equation}
     \frac{P_{p+q}(z)}{z^p} = H_{a b}(z) H_{b a}(z)
\end{equation}
As such, the zeros of $P_{p+q}(z)$ must also be zeros of either $H_{ab}(z)=0$ or $H_{ba}(z)=0$ which is a zero or pole of $M^2(z)$ respectively. 
The zeros of $P_{p+q}(z)$ account for all poles and zeros of $M^2(z)$ at finite and non-zero $z$, however $z=0$ and $z=\infty$ may also be poles and zeros of $M^2(z)$. We are mainly interested if $z=0$ is a pole or zero as we are interested in poles and zeros within the GBZ. Recall that $m>0$ is the highest negative power of $z$ in $H_{ab}(z)$, $n>0$ is the highest negative power of $z$ in $H_{ba}(z)$ and that $\mu = \max\{m,n\}$. At $z=0$, the Laurent polynomials $H_{ab}(z)$ and  $H_{ba}(z)$ are dominated by the $z^{-m}$ and $z^{-n}$ terms respectively and we have
\begin{equation}
    M^2(z)= \frac{H_{a b}(z)}{H_{b a}(z)} \approx  C z^{m-n}
\end{equation}
where $C$ is a constant. If $m=n$, then $M^2(z)$ is finite and $z=0$ is neither a pole nor a zero of $M^2(z)$. If $m>n$, then $z=0$ is a zero of $M^2(z)$ of order $m-n$ and if $n>m$,  then $z=0$ is a pole of $M^2(z)$ of order $n-m$.

We therefore have that the total number of poles and zeros within the GBZ is $p+|m-n|$. Note that $p=m+n$. If $m>n$, the GBZ encloses the first 2$m$ poles and zeros and if $n>m$, the GBZ encloses the first 2$n$ poles and zeros. Hence, the GBZ encloses the first 2$\mu$ poles and zeros of $M^2(z)$. Eq.~\eqref{Eq:WgbzWpz} therefore follows by the argument principle.

\subsection{Edge-state invariant on the $M$-Riemann sphere}
Both the $W_\text{gbz}$ of Eq.~\eqref{Eq:gbz} and the $W_{\text{pz}}$ of Eq.~\eqref{Eq:polezero} are related to the poles and zeros of $M^2(z)$, which are closely related to the ``poles and zeros'' of $M(z)$ (more rigorously, the points where $M(z) \rightarrow \infty$ and $M(z) \rightarrow 0$, respectively). The poles and zeros of $M(z)$ have a simple Riemann-sphere interpretation. The right-eigenvector ratio $M$ is a complex number. We convert $M$ to a Riemann-sphere via the stereographic projection:
\begin{align}
\label{Eq:MRiemannStart}
X(M) &= \frac{2  \Re(M)}{1 + |M|^2}\\ 
Y(M) &= \frac{2  \Im(M)}{1 + |M|^2}\\
Z(M) &= \frac{|M|^2 - 1}{|M|^2 + 1}\label{Eq:MRiemannEnd}
\end{align}
where $(X,Y,Z)$ are Cartesian coordinates of a point on the Riemann sphere. Below we refer to the Riemann sphere as defined by Eqs.~\eqref{Eq:MRiemannStart} to~\eqref{Eq:MRiemannEnd} as the $M$-Riemann sphere.  A point on the $M$-Riemann sphere can be alternatively described in spherical coordinates: i.e. $(X,Y,Z)=(\sin \theta \cos \varphi, \sin \theta \sin \varphi, \cos \theta)$. We can then map the $M$-Riemann sphere to $M$ by 
\begin{equation}
M (\theta, \varphi) = e^{i\varphi} \cot\left(\frac{\theta}{2}\right).\label{Eq:MRiemannSpherical}
\end{equation}
In the maps of Eqs.~\eqref{Eq:MRiemannStart}-\eqref{Eq:MRiemannSpherical}, the poles and zeros of $M(z)$ correspond to the north and south pole on the $M$-Riemann sphere respectively.

Eq.~\eqref{Eq:gbz} describes the the winding of the image of the GBZ about the origin on the complex $M$-plane, which corresponds to the winding number of the image of the GBZ on the $M$-Riemann sphere about the north and south poles. The $M$-Riemann sphere reduces to the typical Bloch sphere for two-level Hermitian models~\cite{asboth2015ashort}, and is equivalent to the non-Hermitian Bloch sphere in Ref.~\cite{lieu208topological} when we take the GBZ as the integration contour.

For simple cases with two-band sublattice symmetry, a subGBZ loop winds around the origin of the $z$-plane once~\cite{wang2024nonhermitian}. For a model in off-diagonal form, the image of such a subGBZ loop that winds twice around the origin on the $M^2(z)$-plane corresponds to a subGBZ loop that winds once around the origin on the $M(z)$-plane. Using the $M^2(z)$ plane is easier for interpreting the pole-zero invariant, as there are no branch cuts, but the winding on $M(z)$ plane (and hence the $M$-Riemann sphere) is more similar to typical Bloch sphere interpretations. A winding number of 2 in Eq.~\eqref{Eq:gbz} corresponds to a subGBZ loop winding around the north-south axis once on the $M$-Riemann sphere which corresponds to two topological edge states in a finite open-boundary case.

\subsection{General sublattice-symmetric edge-state invariant}
\label{sec:gen}
The results of Eq.~\eqref{Eq:gbz} and Eq.~\eqref{Eq:polezero} require the Bloch matrix to be in an off-diagonal form. Here, we show that the results can be generalized to a two-band non-Hermitian Hamiltonian with sublattice symmetry that may not be represented in this form. A two-band Hamiltonian with sublattice symmetry in general is described by Eq.~\eqref{Eq:sublattice} above, where $\Gamma$ may not be equal to $\sigma_z$. However, one can always find a unitarily equivalent model in off-diagonal form for a sublattice symmetric model~\cite{wang2024nonhermitian}. Since $\Gamma$ is both unitary and Hermitian, it is in the form of $\Gamma = \mathbf{d}_\Gamma \cdot \mathbf{\sigma}$, where $\mathbf{d}_\Gamma$ is a unit real vector in three dimensions. As such, there exists a real, orthogonal matrix $R$ that rotates $\mathbf{d}_\Gamma$ to the $Z$-axis in the three dimensional space. Using spherical coordinates, we set $\mathbf{d}_\Gamma = (\sin\theta_\Gamma\cos\phi_\Gamma, \sin\theta_\Gamma\sin\phi_\Gamma, \cos\theta_\Gamma)^T$ with $0 \leq \phi_\Gamma < 2\pi$ and $0 \leq \theta_\Gamma \leq \pi$. Then $R$ can be mapped to a unitary matrix $U$ in the two dimensional Hilbert space of the two-band model where
\begin{equation}
U = \left[\begin{array}{cc}
\cos \theta_\Gamma/2 & -e^{-i\phi_\Gamma} \sin \theta_\Gamma/2\\
e^{i\phi_\Gamma} \sin \theta_\Gamma/2 & \cos \theta_\Gamma/2
\end{array}\right]
\label{Eq:unitary}
\end{equation}
with $\Gamma = U^{-1} \sigma_z U$. As such, the transformed Hamiltonian $U H(z) U^{-1}$ satisfies $\sigma_z (UH(z)U^{-1}) \sigma_z = -UH(z)U^{-1}$ and has the off-diagonal form.

We discuss the generalization of the GBZ invariant in Eq.~\eqref{Eq:gbz} first. For $H(z)$ in the off-diagonal form, the GBZ invariant is the winding of the image of the GBZ on the $M$-Riemann sphere about the $Z$-axis. The $Z$-axis is physically significant in this case because the intersection of the $M$-Riemann sphere with the $Z$-axis corresponds to the $M(z)$ poles and zeros which are the gap closing points~\cite{lieu208topological}. A unitary transform leaves the eigenvalues of a model unchanged but affects the eigenvectors and therefore $M(z)$. It can be shown that such a transform leads to a rotation on the $M$-Riemann sphere that rotates the $Z$-axis to the $\mathbf{d}_\Gamma$-axis (see Supplementary~\ref{sec:unitary}). Thus, the winding number on the $M$-Riemann sphere for a general sublattice symmetric model must be around the $\mathbf{d}_\Gamma$-axis instead of the $Z$-axis. On the other hand, the gap closing points for any two-band non-Hermitian model are the $z$-plane branch points given by~\cite{fu2023anatomy,wang2024nonhermitian}
\begin{equation}
    Q(z)= d_x^2(z)+d_y^2(z)+d_z^2(z)=0.
\end{equation}
More information on why these give the branch points is given in Supplementary~\ref{sec:branch}. The $z$ solutions to the above equation map to one of the two antipodal points on the $M$-Riemann sphere, which we denote as $M_{\text{N}}$ and $M_{\text{S}}$. These two points satisfy $M_{\text{N}}M_{\text{S}}^*=-1$ and are precisely the intersections of the $\mathbf{d}_\Gamma$-axis with the $M$-Riemann sphere. The roots of $Q(z)$ provide an alternative method of deriving the unitary transform required between off-diagonal forms and more general forms.
To use the GBZ form of the general sublattice symmetry invariant, one can find the $M$ value for any one of the branch points on the $M$-Riemann sphere by solving $Q(z) = 0$, and plot the GBZ trajectory on the $M$-Riemann sphere.

We now discuss the generalization of the pole-zero invariant in Eq.~\eqref{Eq:polezero}. For $H(z)$ in the off-diagonal form, the construction of $M^2(z)$ as a single-valued function allows direct identification of its zeros and poles. This construction is no longer suitable for a general sublattice symmetric model as the branch points of $M(z)$, located at $M_{N}$ or $M_{S}$, are generally not coincident with $M = 0$ and $M = \infty$. As such, we will convert the invariant into a form that involves the image of the $z$-plane branch points in $M$, which then generalizes to the cases with general sublattice symmetry.

Consider the pole-zero invariant using $M(z)$ in Eq.~\eqref{Eq:polezero} instead of $M^2(z)$ for a model in off-diagonal form first. For simplicity, let us assume all poles and zeros in $M^2(z)$ have order $\pm 1$. The pole-zero invariant using $M(z)$ then becomes
\begin{equation}
\# (M = 0) - \# (M = \infty)
\label{Eq:polezeroM}
\end{equation}
for the first 2$\mu$ $M(z)$ values where $z$ are the solutions to $z^{2\mu}Q(z) = 0$ sorted by the $|z|$ magnitude. As the higher-order poles or zeros correspond to degenerate $z$ solutions, this formula also applies to higher-order poles or zeros in $M^2$ if we count each $z$ solution exactly once.

Using the fact that a unitary transform leads to a rotation in the $M$-Riemann sphere, the above $M(z)$ pole-zero invariant remains the same for a sublattice-symmetric model not in off-diagonal form, except that we replace $M = 0$ and $M = \infty$ with $M = M_{N}$ or $M = M_{S}$. Depending on which of the $M_{N}$ and $M_{S}$ replaces $0$ and which replaces the other, the $W_{\text{pz}}$ defined will differ by a sign, but the edge state count $|W_{\text{pz}}|$ remains the same.

As such, the pole-zero invariant for general sublattice symmetry using $M(z)$ is
    \begin{equation}
\# (M = M_{N})-\# (M = M_{S})
\label{Eq:polezeroMgeneral}
\end{equation}
for the first 2$\mu$ $M(z)$ values where $z$ are the solutions to $z^{2\mu}Q(z) = 0$ sorted by the $|z|$ magnitude. To calculate the invariant in this form, the first 2$\mu$ solutions to $z^{2\mu}Q(z) = 0$ is listed, possibly including $z=0$. The corresponding $M(z)$ values can be found by substituting $z$ into the definition of $M$ (Eq. \eqref{Eq:Mdef}). Since the $M(z)$ obtained are from $z$ branch points, they can only take two distinct values ($M_{N}$ and $M_{S}$). The invariant then becomes the difference between the number of appearances of $M_{N}$ and $M_{S}$.

\section{Numerical results}
\label{sec:results}

In this section, we illustrate the connections between the GBZ invariant in Eq.~\eqref{Eq:gbz} and pole-zero invariant Eq.~\eqref{Eq:polezero} through numerical examples. The models that we consider include the Hermitian SSH model~\cite{su1979solitons}, a generalized non-Hermitian SSH model with $t_3$ hoppings where $\mu=1$~\cite{guo2021nonhermitian,yao2018edge,kunst2018biorthogonal,yokomizo2019nonbloch}, a longer-range non-Hermitian model with hopping range across three unit cells where $\mu=3$, and an example with winding number in Eq.~\eqref{Eq:gbz} greater than 2 where $\mu=2$. These are all in off-diagonal form and Eq.~\eqref{Eq:gbz} and Eq.~\eqref{Eq:polezero} apply. We then also provide numerical results on the SSH-Creutz model~\cite{zurita2021tunable,mccann2023catalog,bergholtz2021exceptional,longhi2019probing,lee2016anomalous} which has sublattice symmetry but is not in the off-diagonal form, where Eq.~\eqref{Eq:polezeroMgeneral} applies.

For each of these examples, we demonstrate the GBZ invariant in Eq.~\eqref{Eq:gbz} by plotting the image of the GBZ on the $M$-Riemann sphere in subpanel (b) of Figs.~\ref{fig:SSHTop} to~\ref{fig:nhSSHCreutzTriv}. The GBZ invariant is interpreted as the winding around the north-south pole axis for the models in off-diagonal form or the $M_{\text{N,S}}$ axis (i.e the
$\mathbf{d}_{\Gamma} \text {-axis}$) for SSH-Creutz models. 

For the models in off-diagonal form, we also provide plots on the $z$-plane colored by $\arg(M^2(z))=\frac{H_{a b}(z)}{H_{b a}(z)}$ which gives a way to visualize the GBZ invariant in Eq.~\eqref{Eq:Msquared} in subpanel (c) of Figs.~\ref{fig:SSHTop} to~\ref{fig:fouredge}. The color scheme follows domain coloring~\cite{binkowski2024poles} which maps argument values of $\arg(M^2(z))$ from $-\pi$ to $\pi$ as a hue gradient from red through the spectrum back to red. The argument principle then corresponds to the number of hue cycles crossed along a contour on the $z$-plane. For instance, $W=2$ means two hue cycles are crossed along the GBZ path. Poles and zeros are visible as singularities in the color scheme, where the hue cycle wraps in opposite directions around poles when compared to zeros (and vice versa). Using domain colouring, we can also tell what the order of the pole or zeros are, as they correspond to the number of hue cycles emanating from that point. We mark poles and zeros with $\times$ and $\circ$ markers respectively as well as the GBZ path in Figs.~\ref{fig:SSHTop} to~\ref{fig:fouredge}, allowing us to visually count the poles and zeros in the GBZ and verify Eq.~\eqref{Eq:polezero}.

For the SSH-Creutz models which have sublattice symmetry but is not in off-diagonal form, we plot the GBZ values on the $z$-plane as well as N and S markers for the $z$ values that correspond to $M_N$ and $M_S$ points in subpanel (c) for Figs.~\ref{fig:SSHCreutzTop} to~\ref{fig:nhSSHCreutzTriv}, which allows us to verify Eq.~\eqref{Eq:polezeroMgeneral}. We do not color the $z$-plane by $\arg(M^2(z))$ or $\arg(M(z))$ in these plots as these quantities do not have a simple form for general sublattice symmetry form and have branch cuts, making it harder to visually interpret the plots. As the relevant axis is no longer along the north-south pole, the hue cycles crossed along a contour are no longer relevant. 

For Figs.~\ref{fig:SSHTop} to~\ref{fig:SSHCreutzTriv}, in subpanel (a) we show the OBC eigenvalues for $N=100$ unit cells, which leads to 200 OBC eigenvalues as there are two sublattice sites per unit cell. The OBC plot shows whether or not we have topological zero energy edge states. If there are edge states, we also include in subpanel (d) a list plot of the real part of the OBC eigenvalues $\Re(E)$ plotted against their sorted $\Re(E)$ order. This way, the zero energy edge states must appear in the middle of the sorted indices. We include the middle 40 OBC eigenvalues which allows us to visually count the number of degenerate zero energy edge states.

\subsection{Hermitian SSH}

\begin{figure}[h]
\centering
\colorbox{white}{\includegraphics[width=0.5\textwidth]{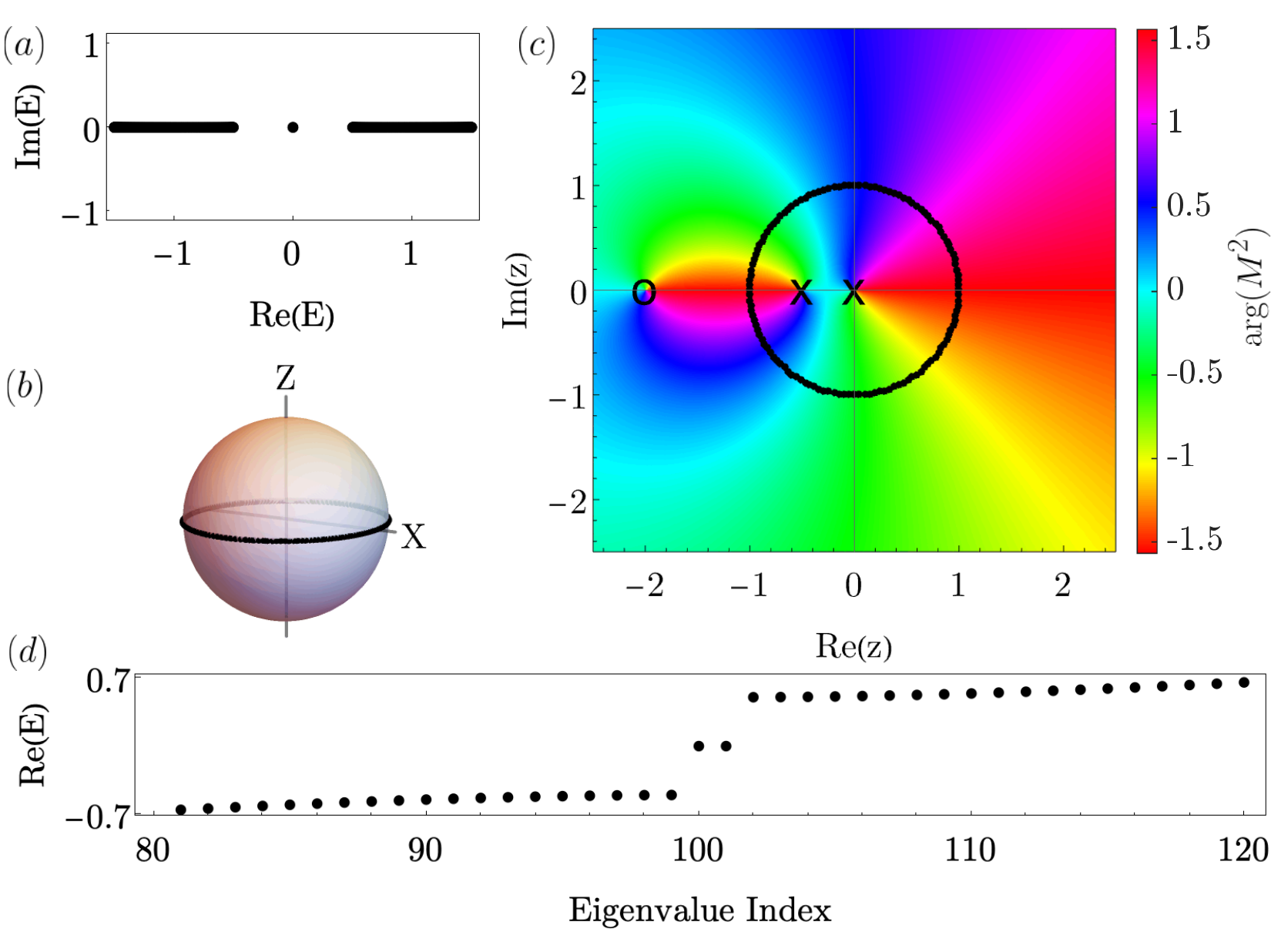}}
\caption{Topological phase of the SSH model with $t_1=0.5, t_2=1$. (a) OBC spectra for $N=100$. (b) The GBZ trajectory on the $M$-Riemann sphere (c) $z$-plane colored by $\arg(M^2(z))$ with GBZ in black scatter, and poles and zeros of $M^2(z)$ as $\times$ and $\circ$ markers respectively. (d) Real part of eigenvalues sorted in increasing order of $\Re(E)$ with sorted eigenvalue index along $x$ axis, to show how many zero energy eigenvalues there are. }
\label{fig:SSHTop}
\end{figure}

We begin with the Hermitian SSH model, which is very well-studied in the literature~\cite{asboth2015ashort,su1979solitons}. It is given by the Hamiltonian 
\begin{equation}
    H_{\text{SSH}}(z)=\left[\begin{array}{cc}
0 & t_1+t_2 / z \\
t_1+t_2 z & 0
\end{array}\right] .
\end{equation}
where $t_1$ are the intracell couplings and $t_2$ are the intercell couplings between the two sublattice sites in a unit cell. In Fig.~\ref{fig:SSHTop}, we use the parameters $t_1=0.5, t_2=1$ for which the SSH model supports topological edge states. In Fig.~\ref{fig:SSHTop}(a), we plot the eigenenergy of this Hamiltonian for a finite lattice with $N = 100$ unit cell sites. We note the presence of edge states at zero energy. For Hermitian models, the GBZ is the same as BZ with $|z| = 1$. In Fig.~\ref{fig:SSHTop}(b), we plot the $M_+(z)$ and $M_-(z)$ contours as $z$ varies on the BZ. Both contours are located at the equator of the $M$-Riemann sphere and winds around the north and the south poles. The results here provides a validation of the connection between the winding on the $M$-Riemann sphere and the existence of topological edge states. In Fig.~\ref{fig:SSHTop}(c), the first $2\mu=2$ poles and zeros of $M^2(z)$ within the GBZ are both poles, which leads to $W=2$ which corresponds to 2 nontrivial topological edge states, as verified in Fig.~\ref{fig:SSHTop}(d). 

\begin{figure}[h]
\centering
\includegraphics[width=0.5\textwidth]{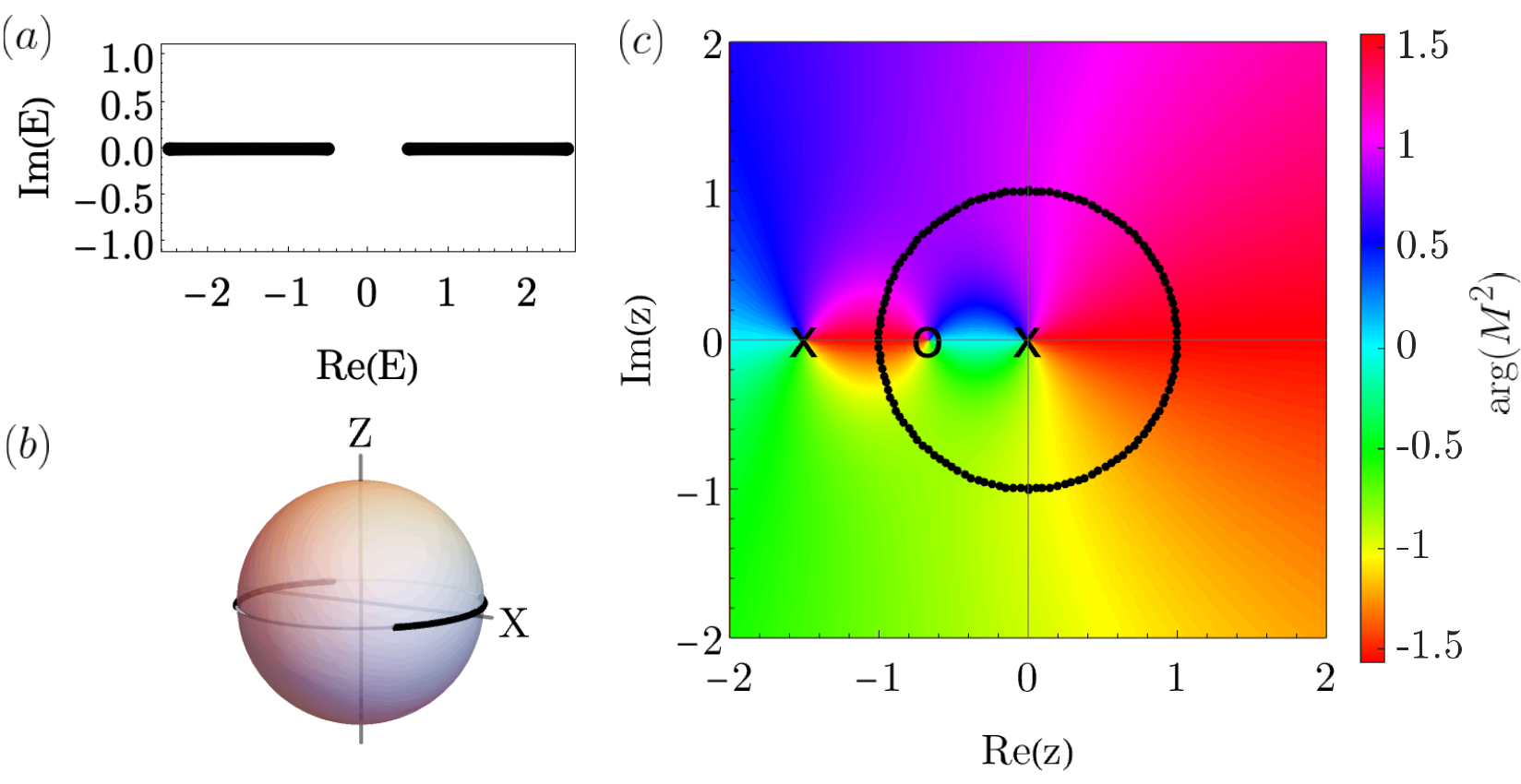}
\caption{Trivial phase of the SSH model with $t_1=1.5, t_2=1$. (a) OBC spectra for $N=100$. (b) The GBZ trajectory on the $M$-Riemann sphere. (c) $z$-plane colored by $\arg(M^2(z))$ with GBZ in black scatter, and poles and zeros of $M^2(z)$ as $\times$ and $\circ$ markers respectively. }
\label{fig:SSHTriv}
\end{figure}
In Fig.~\ref{fig:SSHTriv} we use the parameters  $t_1=1.5, t_2=1$. In Fig.~\ref{fig:SSHTriv}(a), there are no edge states. In Fig.~\ref{fig:SSHTriv}(b) the image of the GBZ on the $M$-Riemann sphere has a trivial winding. In Fig.~\ref{fig:SSHTriv}(c) the first $2\mu=2$ poles and zeros of $M^2(z)$ within the GBZ consist of a pole and a zero, thus $W=0$ and there are no topological zero-energy edge states.

\subsection{Non-Hermitian longer-range SSH models}
\label{sec:nhssh}
We now consider a generalized non-Hermitian version of the SSH model~\cite{guo2021nonhermitian,yao2018edge,kunst2018biorthogonal,yokomizo2019nonbloch} which is in off-diagonal form and has $H_{ab}(z)=\frac{\gamma_1}{2}+t_1-\left(\frac{\gamma_2}{2}-t_2\right)/z+\left(\frac{\gamma_3}{2}+t_3\right)z$ and $H_{ba}(z)=-\frac{\gamma_1}{2}+t_1+\left(\frac{\gamma_2}{2}+t_2\right)z-\left(\frac{\gamma_3}{2}-t_3\right)/z.$ Here, $\mu=1$ as well.

\begin{figure}
\centering
\includegraphics[width=0.5\textwidth]{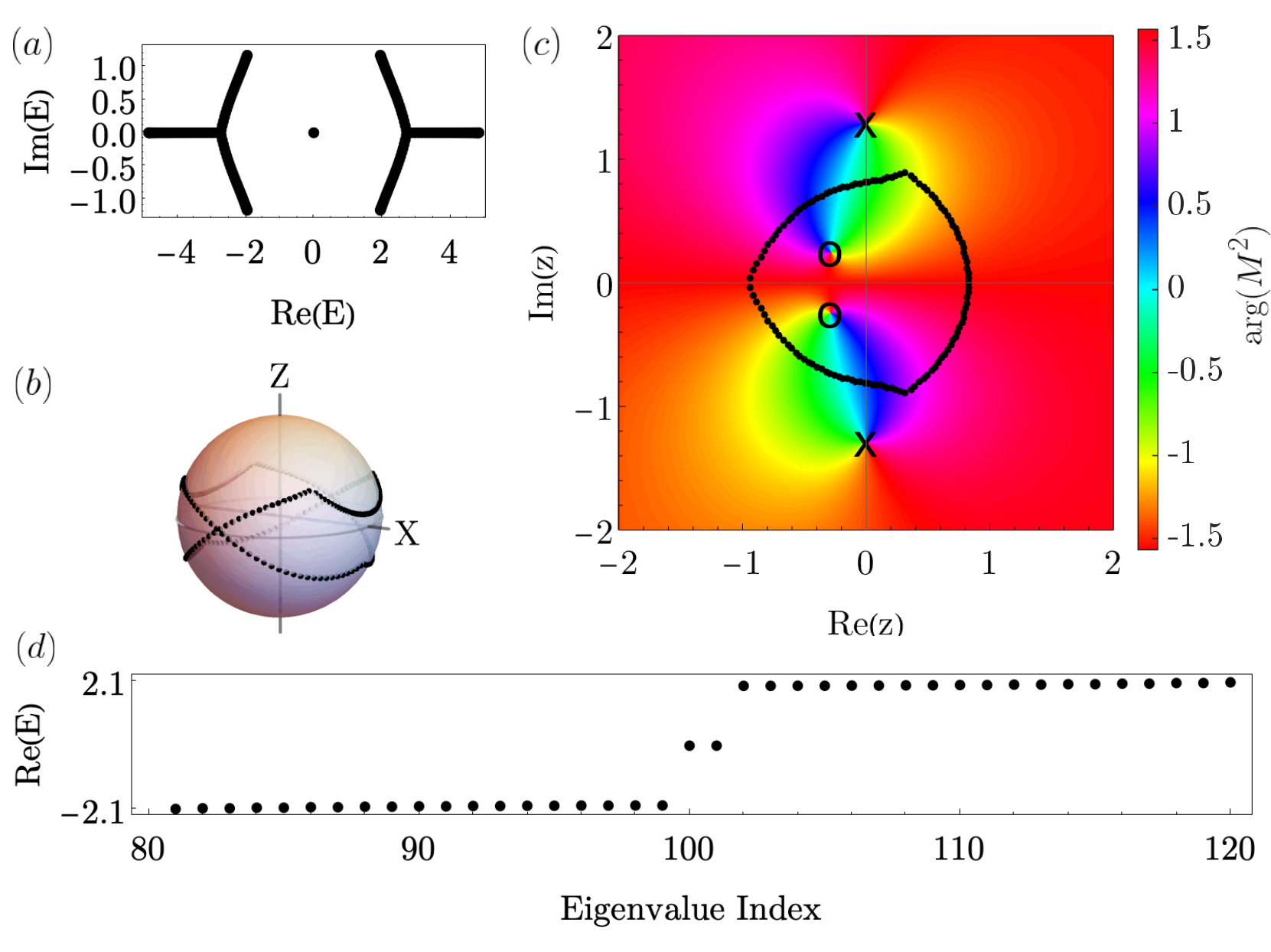}
\caption{Topological phase of the generalized nearest-neighbour non-Hermitian SSH model with $t_1=1, t_2=1, t_3=3, \gamma_1=2, \gamma_2=1, \gamma_3=1$. (a) OBC spectra for $N=100$  (a) OBC spectra for $N=100$. (b) The GBZ trajectory on the $M$-Riemann sphere. (c) $z$-plane colored by $\arg(M^2(z))$ with GBZ in black scatter, and poles and zeros of $M^2(z)$ as $\times$ and $\circ$ markers respectively (d) Real part of eigenvalues sorted in increasing order of $\Re(E)$ with sorted eigenvalue index along $x$ axis, to show how many zero energy eigenvalues there are. }
\label{fig:GenSSHTop}
\end{figure}

In Fig.~\ref{fig:GenSSHTop}, we use the parameters $t_1=1, t_2=1, t_3=3, \gamma_1=2, \gamma_2=1, \gamma_3=1$ for which the system has topological edge states near $E=0$ as seen in Fig.~\ref{fig:GenSSHTop}(a). We see that in Fig.~\ref{fig:GenSSHTop}(b), even though $d_o(z)=d_z(z)=0$, for non-Hermitian models the image of the GBZ on the $M$-Riemann sphere may not be confined to the equator like in Hermitian models. In this case, both subGBZ winds around the north-south pole axis. In Fig.~\ref{fig:GenSSHTop}(c), we show that the first 2 $M^2(z)$ poles and zeros are both zeros, $W=2$ corresponding to two topological edge states, as verified in Fig.~\ref{fig:GenSSHTop}(d).

\begin{figure}
\centering
\includegraphics[width=0.5\textwidth]{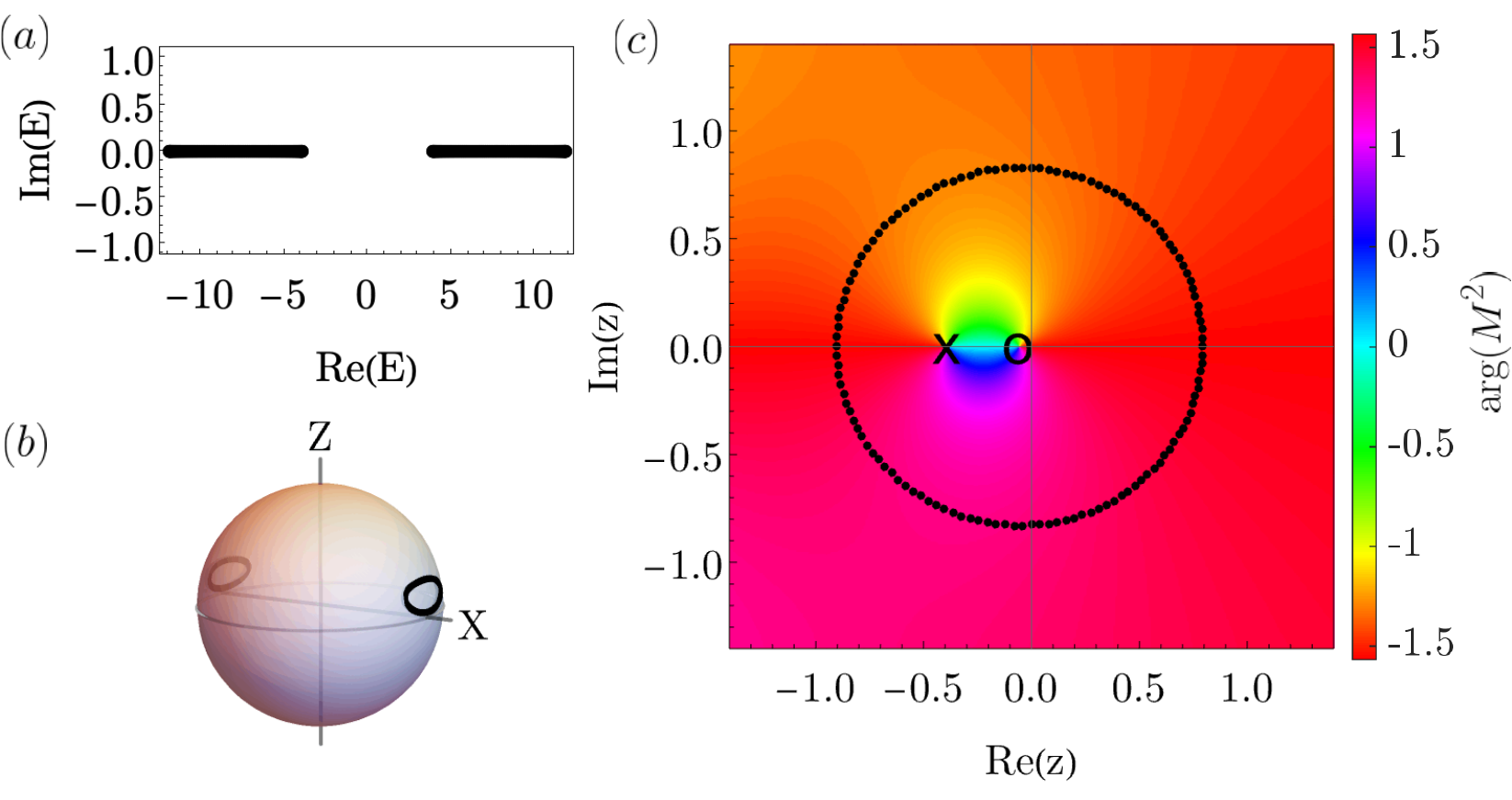}
\caption{Trivial phase of the generalized non-Hermitian SSH model with $t_1=8, t_2=1, t_3=3, \gamma_1=2, \gamma_2=1 \gamma_3=1$.  (a) OBC spectra for $N=100$. (b) The GBZ trajectory on the $M$-Riemann sphere. (c) $z$-plane colored by $\arg(M^2(z))$ with GBZ in black scatter, and poles and zeros of $M^2(z)$ as $\times$ and $\circ$ markers respectively. }
\label{fig:GenSSHTriv}
\end{figure}

In Fig.~\ref{fig:GenSSHTriv}, we use the parameters $t_1=8, t_2=1, t_3=3, \gamma_1=2 ,\gamma_2=1, \gamma_3=1$ for which the system has no topological edge states, as seen in Fig.~\ref{fig:GenSSHTriv}(a). We see that in Fig.~\ref{fig:GenSSHTriv}(b), the image of both subGBZ loops do not wind around the north-south pole axis. In Fig.~\ref{fig:GenSSHTriv}(c), we show that the first 2 $M^2(z)$ poles and zeros are a zero and a pole, thus $W=0$ indicating no topological edge states.

\begin{figure}
\centering
\includegraphics[width=0.5\textwidth]{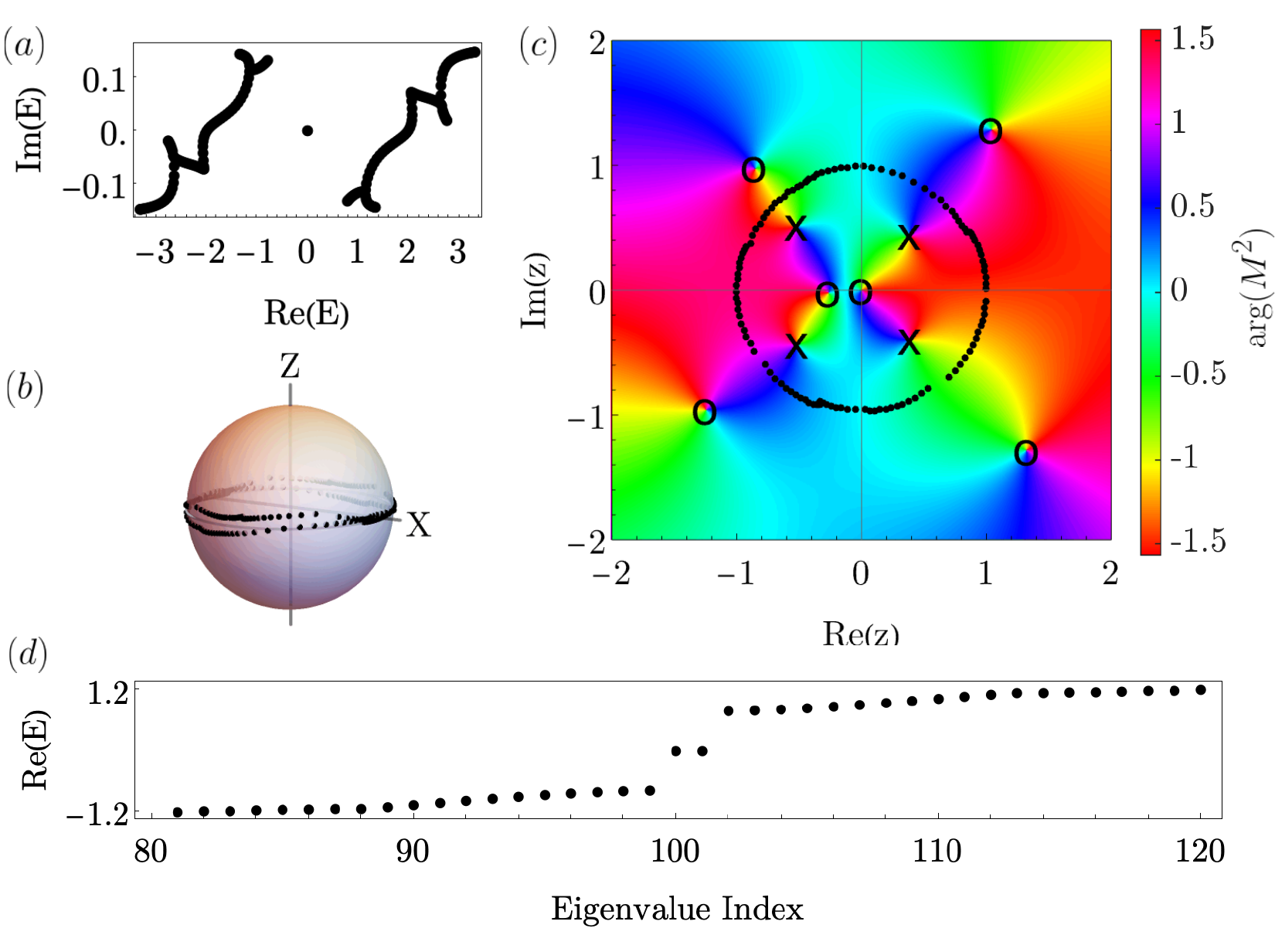}
\caption{Topological phase of the longer-range SSH with $t_1 = 0.5+0.5  i, t^\prime_1 = 2 , t_2 = 0.5 -0.2 i, t^\prime_2 = 2 , v = 0.5, u = 0.3$.   (a) OBC spectra for $N=100$. (b) The GBZ trajectory on the $M$-Riemann sphere. (c) $z$-plane colored by $\arg(M^2(z))$ with GBZ in black scatter, and poles and zeros of $M^2(z)$ as $\times$ and $\circ$ markers respectively. (d) Real part of eigenvalues sorted in increasing order of $\Re(E)$ with sorted eigenvalue index along $x$ axis, to show how many zero energy eigenvalues there are. }
\label{fig:NNNTop}
\end{figure}

We now consider an even longer range extension which is also in off-diagonal form and has $H_{a b}(z)=t_1+t_1^{\prime} / z+v / z^2+u z^3, H_{b a}(z)=t_2+t_2^{\prime} z+v z^2+u / z^3 .$ This time, $\mu=3$ and we are interested in the first 6 $M(z)$ pole and zeros.

In Fig.~\ref{fig:NNNTop}, we use the parameters $t_1 = 0.5+0.5  i, t^\prime_1 = 2 , t_2 = 0.5 -0.2 i, t^\prime_2 = 2 , v = 0.5, u = 0.3$ for which the system has topological edge states near $E=0$, as seen in Fig.~\ref{fig:NNNTop}(a). We see that in Fig.~\ref{fig:NNNTop}(b) the image of both subGBZ loops wind around the north-south pole axis. In Fig.~\ref{fig:NNNTop}(c), we show that the first 6 $M(z)$ poles and zeros consists of four poles and two zeros, $W=2$ indicating two topological edge states, as verified in Fig.~\ref{fig:NNNTop}(d).
  
\begin{figure}
\centering
\includegraphics[width=0.5\textwidth]{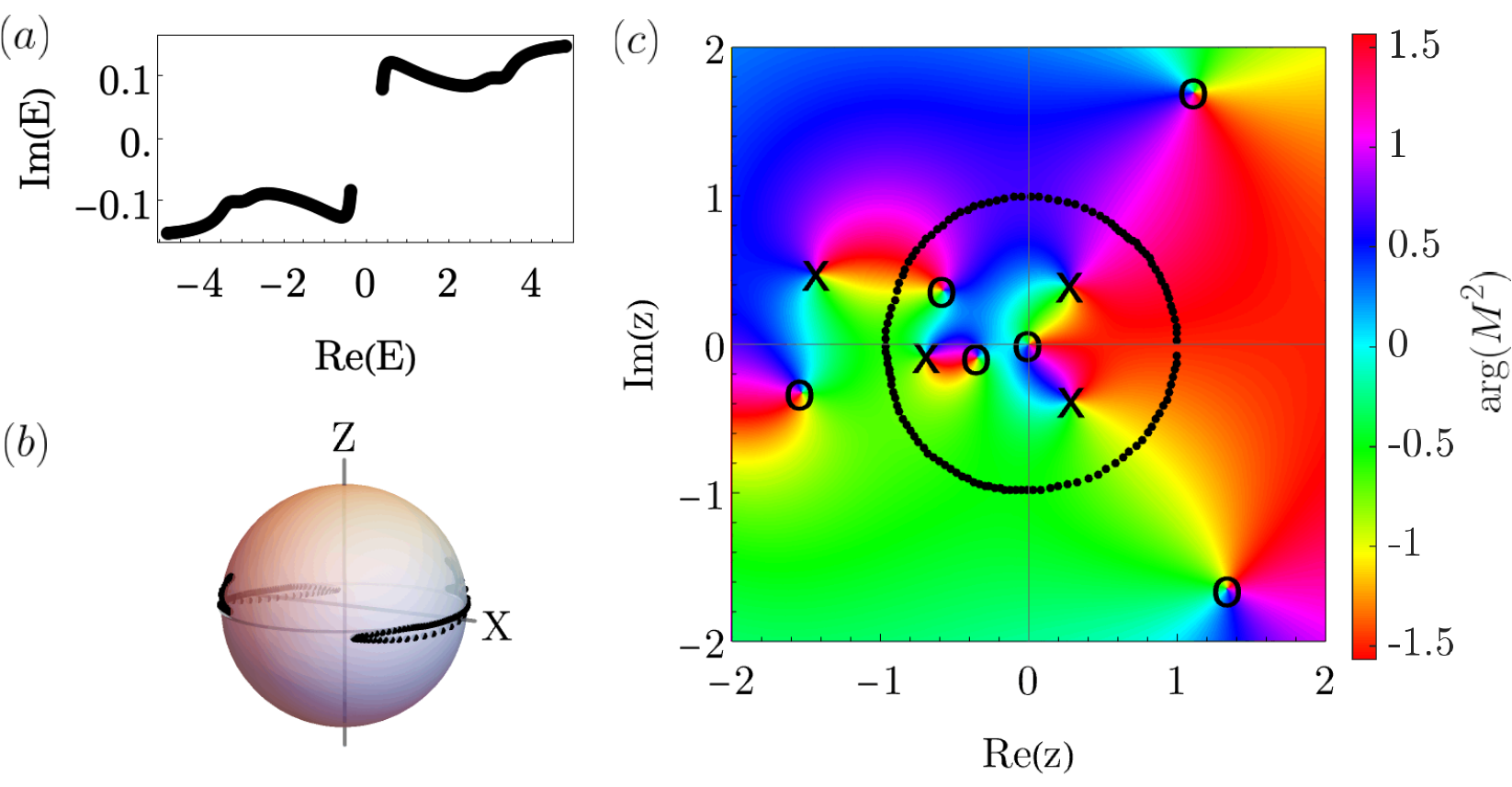}
 \caption{Trivial phase of the longer-range SSH with $t_1 = 2+0.5 i, t^\prime_1 = 2 , t_2 = 2 -0.2 i, t^\prime_2 = 2 , v = 0.5, u = 0.3$.  (a) OBC spectra for $N=100$. (b) The GBZ trajectory on the $M$-Riemann sphere. (c) $z$-plane colored by $\arg(M^2(z))$ with GBZ in black scatter, and poles and zeros of $M^2(z)$ as $\times$ and $\circ$ markers respectively.  }
\label{fig:NNNTriv}
\end{figure}

In Fig.~\ref{fig:NNNTriv}, we use the parameters $t_1 = 2+0.5 i, t^\prime_1 = 2 , t_2 = 2 -0.2 i, t^\prime_2 = 2 , v = 0.5, u = 0.3$ for which the system has no topological edge states, as seen in Fig.~\ref{fig:NNNTriv}(a). We see that in Fig.~\ref{fig:NNNTriv}(b), both subGBZ loops do not wind around the north-south pole axis. In Fig.~\ref{fig:NNNTriv}(c), we show that the first 6 $M(z)$ poles and zeros consists of three poles and three zeros, thus $W=0$ indicating no topological edge states.

\subsection{Example with winding number larger than 2}
We now consider an example that winding numbers in Eq.~\eqref{Eq:gbz} larger than 2 which corresponds to more than two edge states. Consider a non-Hermitian generalization of a model in Ref.~\cite{chen2021connection} with $H_{ab}(z)= t_0+\frac{t_1+g_1}{z}+\frac{t_2+g_2}{z^2}$ and $H_{ba}(z)= t_0 +(t_1-g_1)z+(t_2-g_2)z^2$ with $t_0 =5 ,t_1 = 10 , t_2 = 15 ,  g_1 =1 , g_2 = 1$.
\begin{figure}
\centering
\includegraphics[width=0.5\textwidth]{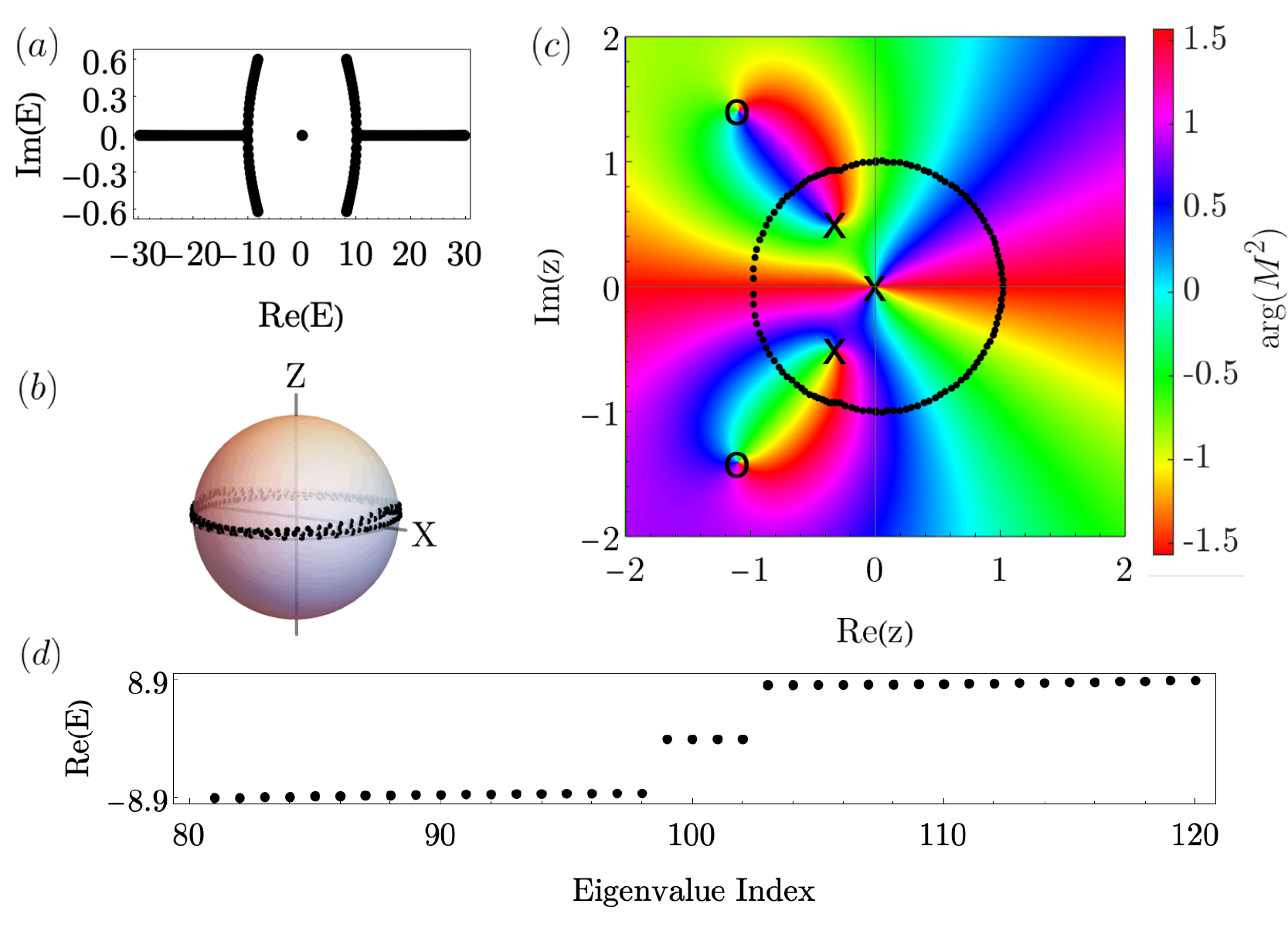}
\caption{Topological phase of a model with winding number larger than 2  with $t_0 =5 ,t_1 = 10 , t_2 = 15 , g_1 =1 , g_2 = 1$. (a) OBC spectra for $N=100$. (b) The GBZ trajectory on the $M$-Riemann sphere. (c) $z$-plane colored by $\arg(M^2(z))$ with GBZ in black scatter, and poles and zeros of $M^2(z)$ as $\times$ and $\circ$ markers respectively. Here the pole at $z=0$ has order 2. (d) Real part of eigenvalues sorted in increasing order of $\Re(E)$ with sorted eigenvalue index along $x$ axis, to show how many zero energy eigenvalues there are. }
\label{fig:fouredge}
\end{figure}

In Fig.~\ref{fig:fouredge}(a), we see that there are topological edge states. In Fig.~\ref{fig:fouredge}(b), we see that both subGBZ loops wind around the north-south pole axis twice. In Fig.~\ref{fig:fouredge}(c), noting that the pole at the origin has order two (as there are two hue cycles emanating from that point) and that the first $2\mu=4$ poles and zeros of $M^2(z)$ consist of four poles, we have $W=4$ corresponding to four edge states, as verified in Fig.~\ref{fig:fouredge}(d).

\subsection{SSH-Creutz model}

We now consider a sublattice symmetric model not in the off-diagonal form. To do this, we start with the Creutz model~\cite{longhi2019probing,lee2016anomalous,creutz1999end} given by:
\begin{equation}
    H_{\text{Creutz}}(z)=\left[\begin{array}{cc}
i t_2 /(2 z)-i t_2 z / 2 & t_1+t_2 /(2 z)+t_2 z / 2 \\
t_1+t_2 /(2 z)+t_2 z / 2 & -i t_2 /(2 z)+i t_2 z / 2
\end{array}\right]
\end{equation}
This model has the same eigenvalues as the SSH model and is equivalent under a unitary transform~\cite{bergholtz2021exceptional}. We will consider the following SSH-Creutz model described by 
\begin{equation}
    H(z)=a H_{\mathrm{SSH}}(z)+b H_{\text{Creutz}}(z)
\end{equation}
where $a$ and $b$ can be considered as weightings of the SSH and Creutz model respectively. Here, $\mu=1$ and we are interested in the first 2 $M_{\text{N}}$ and $M_{\text{S}}$ points as discussed in Sec.~\ref{sec:gen}.

\begin{figure}
\centering
\includegraphics[width=0.5\textwidth]{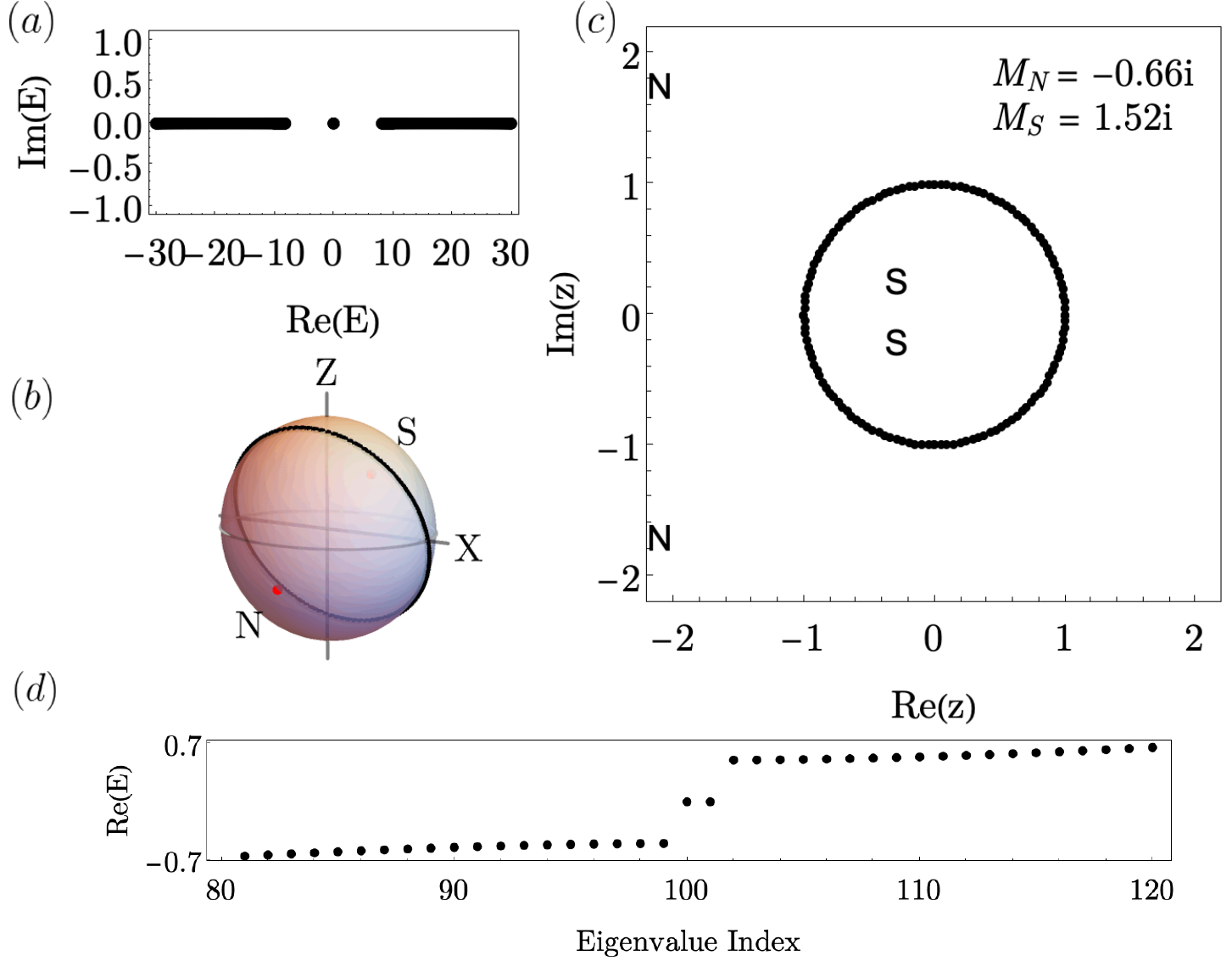}
\caption{Topological phase of SSH-Creutz model with $a = 0.3,b=0.7, t_1 = 0.5, t_2 = 1$ (a) OBC spectra for $N=100$. (b) The GBZ trajectory in black and image of $z$-plane branch point as red dot on $M$-Riemann sphere. (c) $z$-plane with $M_\text{N}$ and $M_{\text{S}}$ marked as N and S markers respectively. (d) Real part of eigenvalues sorted in increasing order of $\Re(E)$ with sorted eigenvalue index along $x$ axis, to show how many zero energy eigenvalues there are. }
\label{fig:SSHCreutzTop}
\end{figure}

In Fig.~\ref{fig:SSHCreutzTop}, we use the parameters $a = 0.3,b=0.7, t_1 = 0.5, t_2 = 1$  for which the system has topological edge states, as seen in Fig.~\ref{fig:SSHCreutzTop}(a). In Fig.~\ref{fig:SSHCreutzTop}(b), we also plot the image of the $z$-plane branch points on the $M$-Riemann sphere as a red dots. They map to the antipodal points $M(z)= -0.66i$ or $1.52 i$, which in $M$-Riemann sphere coordinates is $(X(M),  Y(M), Z(M)) =$ $(0, 0.92, 0.39)$, $(0, -0.92, -0.39))$. Call these two points N and S respectively. We see that both subGBZ wind around the axis formed by these two antipodal points, indicating a topological phase. In Fig.~\ref{fig:SSHCreutzTop}(c), the $M_{\text{N}}$ and $M_{\text{S}}$ are marked as N and S on the $z$-plane. In this case, the first $2\mu=2$ $M_{\text{N}}$ or $M_{\text{S}}$ points inside the GBZ are $M(z)=1.52 i$ and $1.52i$, (or $(X(M),  Y(M), Z(M)) =$ $(0, 0.92, 0.39),(0, 0.92, 0.39)$). Both of these $M_{\text{N}}$ or $M_{\text{S}}$ points inside the GBZ are S poles, which indicates a topological phase.

\begin{figure}
\centering
\includegraphics[width=0.5\textwidth]{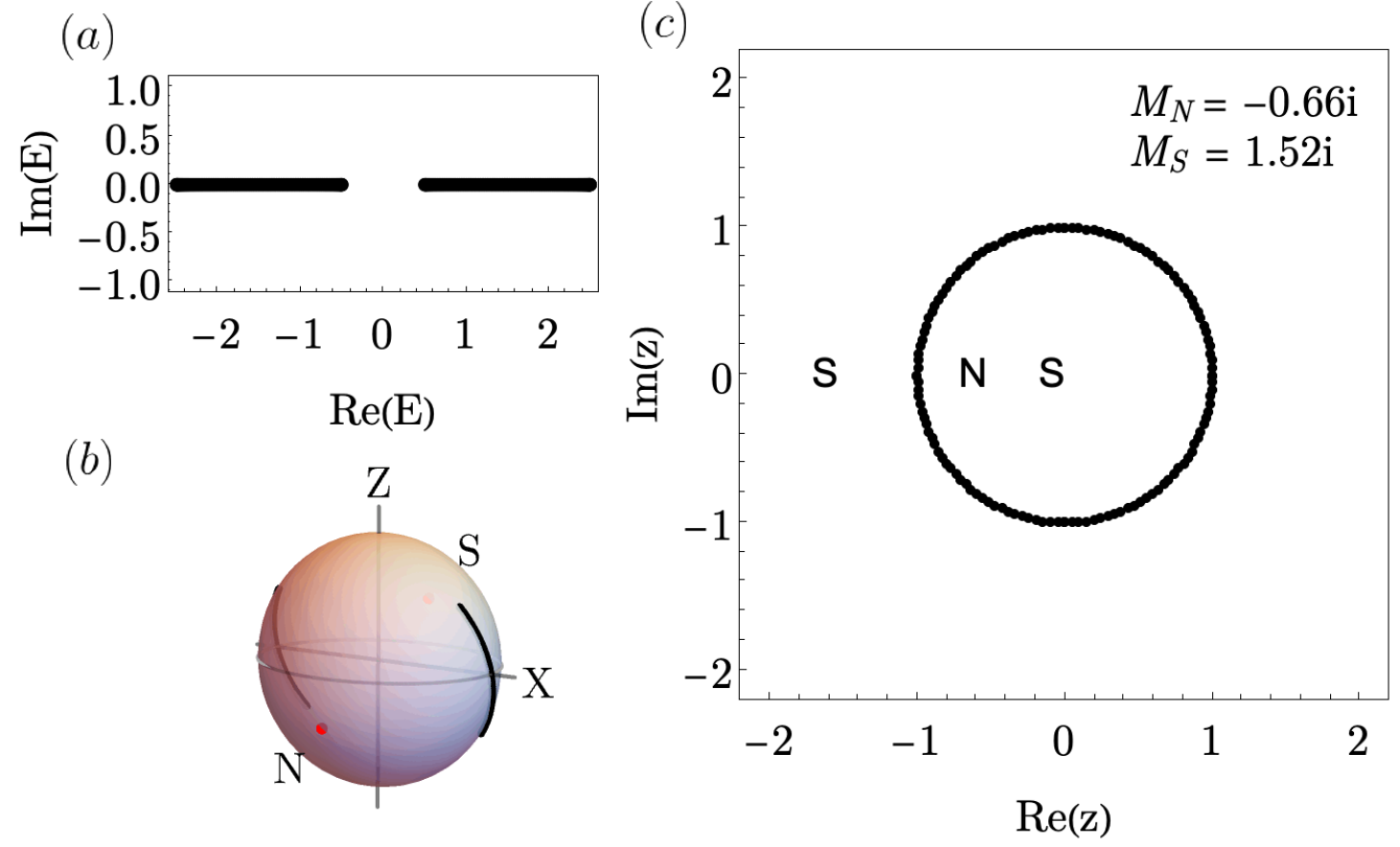}
\caption{Trivial phase of SSH-Creutz model with $a = 0.3,b=0.7, t_1 = 1.5, t_2 = 1$ (a) OBC spectra for $N=100$. (b) The GBZ trajectory in black and image of $z$-plane branch point as red dot on $M$-Riemann sphere. (c) $z$-plane with $M_\text{N}$ and $M_{\text{S}}$ marked as N and S markers respectively. }
\label{fig:SSHCreutzTriv}
\end{figure}

In Fig.~\ref{fig:SSHCreutzTriv}, we use the parameters $a = 0.3,b=0.7, t_1 = 1.5, t_2 = 1$ which has no edge states, as seen in Fig.~\ref{fig:SSHCreutzTriv}(a). In Fig.~\ref{fig:SSHCreutzTriv}(b), we again we see that both subGBZ do not wind around the axis formed by these two antipodal points, indicating a topological phase. In Fig.~\ref{fig:SSHCreutzTop}(c), the first $2\mu=2$ $M_{\text{N,S}}$ points inside the GBZ are $M(z)=-0.66i$ and $1.52 i$, (or $(X(M),  Y(M), Z(M)) =$ $(0, 0.92, 0.39),(0, 0.92, -0.39)$).  These $M_{\text{N}}$ or $M_{\text{S}}$ points consist of a S and an N pole, which indicates that the system is in a trivial phase.
\subsection{Non-Hermitian SSH-Creutz model}
Let us consider the non-Hermitian Creutz model~\cite{lee2016anomalous,bergholtz2021exceptional}:
    \begin{equation}
H_{\text{nhCreutz}}(z)=\left[\begin{array}{cc}
\frac{-i\left(t_2+\gamma\right)}{2 z}+\frac{i t_2 z}{2} & t_1+\frac{t_2}{2 z}+\frac{t_2 z}{2} \\
t_1+\frac{t_2}{2 z}+\frac{t_2 z}{2} & \frac{i\left(t_2+\gamma\right)}{2 z}-\frac{i t_2 z}{2}
\end{array}\right]
\end{equation}
as well as the generalized non-Hermitian SSH described above (and call it $H_{\text{nhSSH}}(z)$) but where $\gamma_1=\gamma=0.7, \gamma_2=\gamma_3=t_3=0$. Now consider the following non-Hermitian SSH-Creutz model:
\begin{equation}
H(z)=a H_{\text {nhSSH }}(z)+b H_{\text {nhCreutz }}(z)
\end{equation}
which is non-Hermitian, has sublattice symmetry and is not in off-diagonal form.
\begin{figure}
\centering
\includegraphics[width=0.5\textwidth]{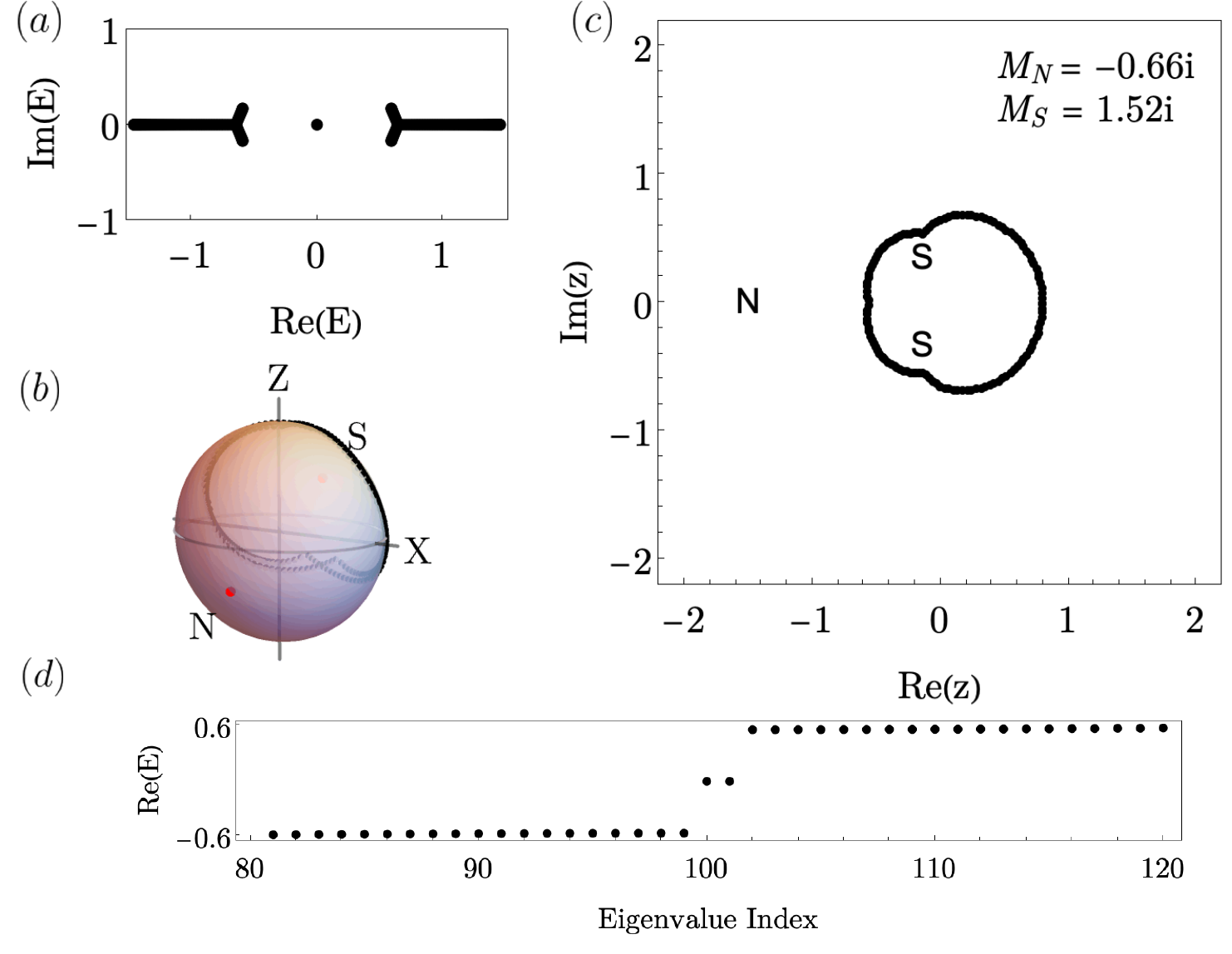}
\caption{Topological phase of non-Hermitian SSH-Creutz model with $a = 0.3,b=0.7, t_1 = 0.5, t_2 = 1,\gamma=0.7$ (a) OBC spectra for $N=100$. (b) The GBZ trajectory in black and image of $z$-plane branch point as red dot on $M$-Riemann sphere. (c) $z$-plane with $M_\text{N}$ and $M_{\text{S}}$ marked as N and S markers respectively.}
\label{fig:nhSSHCreutzTop}
\end{figure}

In Fig.~\ref{fig:nhSSHCreutzTop}, we use the parameters $a = 0.3,b=0.7, t_1 = 0.5, t_2 = 1$  which has topological edge states. We make the same conclusions as in Fig.~\ref{fig:SSHCreutzTop} except that in Fig.~\ref{fig:nhSSHCreutzTop} we have a non-Hermitian example. The main difference is that the GBZ is not confined to $|z|=1$, and the GBZ trajectory on the $M$-Riemann sphere does not have to be on the equatorial plane perpendicular to the $M_{\text{N,S}}$ axis due to non-Hermiticity.

\begin{figure}
\centering
\includegraphics[width=0.5\textwidth]{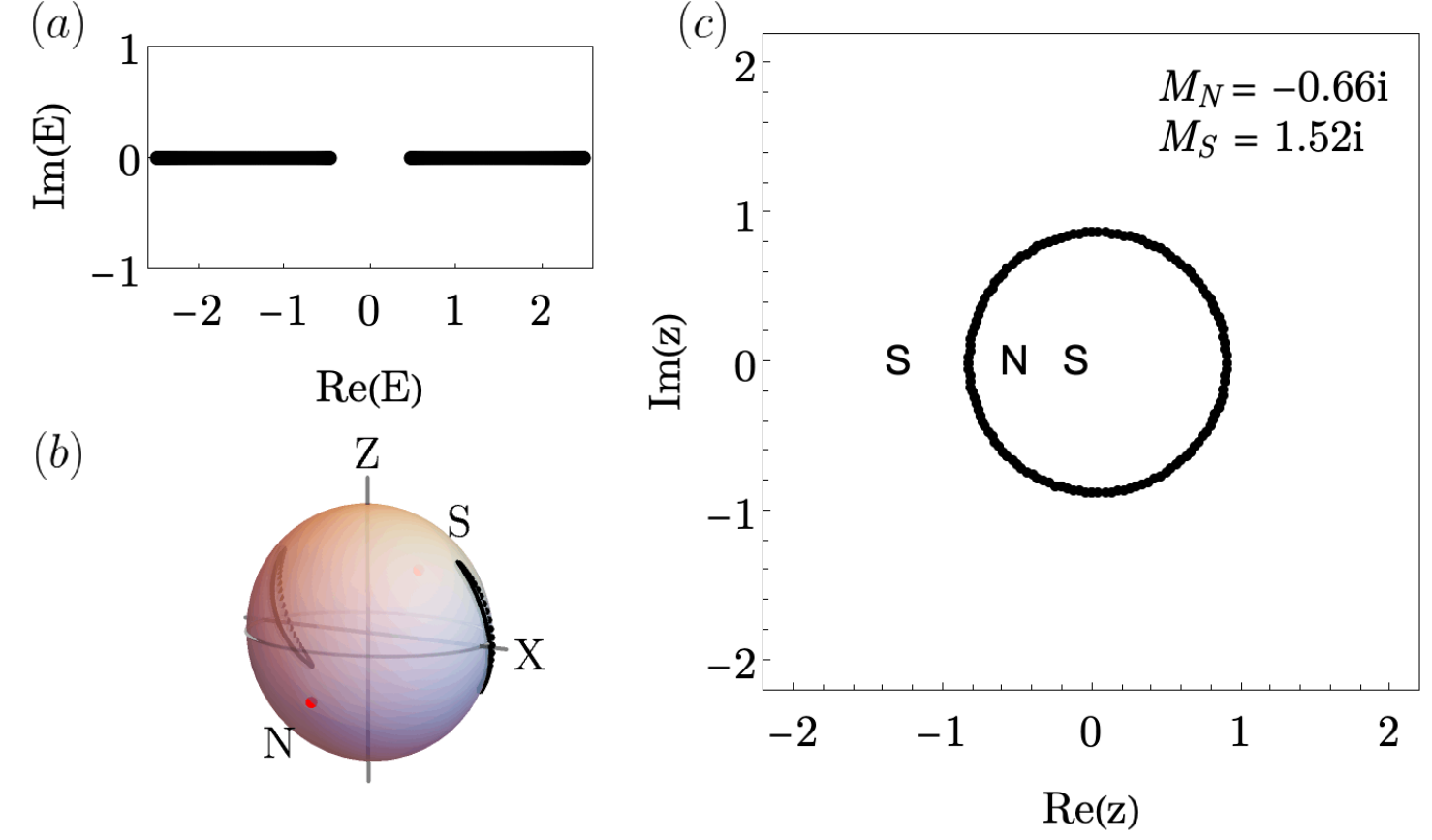}
\caption{Trivial phase of non-Hermitian SSH-Creutz model with $a = 0.3,b=0.7, t_1 = 1.5, t_2 = 1,\gamma=0.7$ (a) OBC spectra for $N=100$. (b) The GBZ trajectory in black and image of $z$-plane branch point as red dot on $M$-Riemann sphere. (c) $z$-plane with $M_\text{N}$ and $M_{\text{S}}$ marked as N and S markers respectively.}
\label{fig:nhSSHCreutzTriv}
\end{figure}

In Fig.~\ref{fig:nhSSHCreutzTriv}, we use the parameters $a = 0.3,b=0.7, t_1 = 1.5, t_2 = 1$  which has no topological edge states. We make the same conclusions as Fig.~\ref{fig:SSHCreutzTriv} but for a non-Hermitian example. Thus, we see that our generalized sublattice symmetry GBZ and pole-zero invariant also applies in the case of non-Hermiticity.

\vspace{2em}
\section{Conclusion}
\label{sec:conclusion}
In conclusion, we have demonstrated that for non-Hermitian, two-band, sublattice-symmetric tight-binding models in off-diagonal form, the GBZ invariant in Eq.~\eqref{Eq:gbz} is equivalent to the pole-zero invariant in Eq.~\eqref{Eq:polezero}. By introducing an $M$-Riemann sphere interpretation of the GBZ invariant based on the eigenvector ratio $M$, we derived a more generalized version of the pole-zero invariant applicable to sublattice-symmetric models not necessarily in off-diagonal form. This $M$-Riemann sphere interpretation of the GBZ invariant also extends to models beyond the off-diagonal form. We numerically illustrated these invariants and cases using non-Hermitian SSH models, their variants, and non-Hermitian SSH-Creutz models. In many cases for sublattice symmetric models, the calculation of the pole-zero invariant can be numerically or analytically easier than the GBZ invariant as it is a lower order polynomial equation. By establishing the equivalence of these approaches, we hope that the tools from both methods can be used more interchangeably or provide more insights for either method. We also hope that the $M$-Riemann sphere formalism introduced in this paper may yield further insights in non-Hermitian systems beyond this symmetry class or that connections to concepts in this work may be found in non-tight-binding models~\cite{felbacq2023characterizing}. A deeper understanding of the topological invariants for 1D non-Hermitian sublattice symmetric edge states~\cite{hou2023topological,longhi2019probing,kunst2018biorthogonal,lee2016anomalous,yao2018edge,yang2020nonhermitian,yokomizo2019nonbloch,jin2019bulk} may lead to new insights in the interplay of edge states with other phenomena (such as lasing~\cite{longhi2018nonhermitian,zhu2022anomalous}, squeezed states~\cite{wan2023quantum,slim2024optomechanical,mcdonald2018phase,busnaina2023quantum}, subsymmetry protected systems~\cite{verma2024non}, the critical non-Hermitian skin effect~\cite{li2020critical}, braiding of edge states~\cite{zhu2024versatile} and more) or models beyond this work (such as multiband~\cite{liu2023topological,nehra2022topology}, driven~\cite{vyas2021topological,wu2020floquet,cao2021non}, higher dimensional~\cite{yang2024anatomy,zhu2024brief,edvardsson2019non,yao2018chern}, nonlinear~\cite{hang2021nonlinear}, stochastics~\cite{tang2021topology}, continuous models~\cite{felbacq2023characterizing,hu2024non,yokomizo2024non} and more).

\begin{acknowledgements}
J.Z acknowledges discussions on related concepts for an earlier version of this work with C. Wojcik and members/ alumni of the Z. Wang group. She also thanks C. Lee and D. Felbacq for email correspondence. This work is funded by a
Simons Investigator in Physics grant from the Simons Foundation (Grant No. 827065), and by  a MURI
project from the U.S. Air Force of Office of Scientific Research (Grant No
FA9550-21-1-0244). J.Z was supported by a Fulbright Future Scholarship and the Quad Fellowship.
\end{acknowledgements}

\section{Appendix}

\subsection{Calculating $z$-plane branch points}
\label{sec:branch}
The zeros of $P_{p+q}(z)$ is related to the $z$-plane branch points of Eq.~\eqref{Eq:char}. To determine these branch points, we rewrite the characteristic polynomial as a polynomial in $E$ and $z$:
\begin{equation}
    E^2 z^p-P_{p+q}(z)=0
\end{equation}
The nonzero $z$-plane branch points occur when the discriminant of the above equation with respect to $E$ is zero~\cite{wang2024nonhermitian,wang2024onedimensional}, which corresponds to the roots of $P_{p+q}(z)$. This can alternatively be seen by noting that one can write $E(z)=\sqrt{Q(z)}=\sqrt{P_{p+q}(z)/z^p}$. Branch points are given when the term in the radical is zero, i.e. when $P_{p+q}(z)=0$. Thus, the $z$-plane branch points occur at $E=0$ for sublattice symmetric models, which are the gap closing points~\cite{fu2023anatomy}.

In addition the $z$-plane branch points as determined by $P_{p+q}(z) = 0$, the points $z=0, \infty$ may also be branch points~\cite{wang2024nonhermitian,wang2024onedimensional}. To check whether $z=0$ is a branch point, we consider
\begin{equation}
     Q(z) = \sum_{l=-p}^q c_l z^l.
\end{equation}
At $z=0$, $Q(z)$ is dominated by the $z^{-p}$ term. Here, $c_l$ are constants. Then $E(z)$ around $z=0$ is
\begin{equation}
    E \approx \sqrt{c_{-p} z^{-p}}=\sqrt{c_{-p}} \cdot z^{-p / 2}.
\end{equation}
As such, $z = 0$ is a branch point if and only if $p$ is odd~\cite{cavalieri2016riemann}. Similarly, $z=\infty$ is a branch point if and only if $q$ is odd.

Note that when there is a branch point at $z=0$ or $z=\infty$ for a model in off-diagonal form, it necessarily corresponds to a pole or zero of $M(z)$. However, the converse is not true, and one can have a pole and zero of $M(z)$ without being a branch point. To see this, note that near $z=0$, 
\begin{align}
    &E(z) \approx \sqrt{c_{-p}} z^{-p / 2} \\
    &H_{ab}(z) \approx a_{-m} z^{-m } \\
    &H_{ba}(z) \approx b_{-n} z^{-n} 
\end{align}
where $c_{-p}, a_{-m},b_{-n} $ are constants. Since $p=m+n$, if $p$ is even, and $m \neq n$, then $z=0$ is a pole or zero of $M(z)$ but it is not a $z$-plane branch point.

\subsection{Eq.~\eqref{Eq:polezero} implies the statement in Ref.~\cite{lee2019anatomy}}
\label{sec:leepolezero}

We first restate the theorem from Ref.~\cite{lee2019anatomy} for the existence of topological edge modes using the notation from our paper. The statement is that an isolated topological zero mode exists when the $m' + n'$ members from the roots of $H_{ab}H_{ba}$ having the largest magnitude do not contain $m'$ members from the roots of $H_{ab}$ and $n'$ members from the roots of $H_{ba}$. Here, $m'$ ($n'$) is the highest positive power of $z$ in $H_{ab}(z)$ ($H_{ba}(z)$).

We first note that Ref.~\cite{lee2019anatomy} uses the largest roots. By taking the complement of the set of roots, we get an equivalent formulation of this statement: edge modes exist when the $m + n$ members from the roots of $H_{ab}H_{ba}$ having the smallest magnitude do not contain $m$ members from the roots of $H_{ab}$ and $n$ members from the roots of $H_{ba}$. The two statements have a similar form because reversing the entire tight-binding chain is equivalent to $z \rightarrow 1/z$, which takes the largest roots to the smallest roots while preserving the spectrum.

In Ref.~\cite{lee2019anatomy}, $z=0$ and $z=\infty$ are sometimes included as roots, but it is not specified when this is needed. To compare with our statement, it becomes necessary to include these roots when $m \neq n$. If $m > n$, then we add $(m-n)$ counts of $z=0$ roots to $H_{ba}(z)$, and vice versa. The extra $z=0$ roots are guaranteed to rank first, and the statement becomes that the first $2\mu$ roots are not $\mu$ roots from $H_{ab}(z)$ and $\mu$ roots from $H_{ba}(z)$.

Finally, note that the set of roots, including $z=0$ and counted with multiplicity, can be succinctly written as the roots from the polynomial equation $z^{|m-n|}(z^mH_{ab})(z^nH_{ba}) = 0$. This simplifies to $z^{2\mu}Q(z) = 0$, and lists all of the poles and zeros of $M^2(z)$ in the off-diagonal basis. As the roots from $H_{ab}(z)$ and the roots from $H_{ba}(z)$ are the poles and zeros of $M^2(z)$ respectively, the statement becomes that there must be an unequal number of poles and zeros of $M^2(z)$ corresponding to the first $2\mu$ roots for edge mode existence, which is a special case of Eq.~\eqref{Eq:polezero} with $W_{\text{pz}} \neq 0$.
\vspace{5 pt}
\subsection{Unitary transformation in $H$ and rotation of the $M$-Riemann sphere} 
\label{sec:unitary}
Let $U$ be a unitary matrix and $H$ be a $2\times 2$ Bloch matrix. Assume $H \ket{\psi} = E \ket{\psi}$. A unitary transformation on $H$ gives $H' = U H U^{-1}$ where $H'$ has the same eigenvalues $E$ as $H$ but transformed eigenvectors $U \ket{\psi}$:
\begin{align}
    \left( U H U^{-1} \right) \left( U \ket{\psi} \right)  &= U H \ket{\psi}\\
    &= U E \ket{\psi} = E \left( U \ket{\psi} \right).
\end{align}
An arbitrary unitary matrix can be written in the form:
\begin{equation}
   U =  \begin{bmatrix}
e^{i \alpha} \cos \theta & e^{i \beta} \sin \theta \\
-e^{-i \beta} \sin \theta & e^{-i \alpha} \cos \theta
\end{bmatrix}
\end{equation}
where $\alpha, \beta, \theta \in \mathbb{R}$. Consider
\begin{align}
   U \left[\begin{array}{c}
v_1^R \\
v_2^R
\end{array}\right] &=  \begin{bmatrix}
e^{i \alpha} \cos \theta & e^{i \beta} \sin \theta \\
-e^{-i \beta} \sin \theta & e^{-i \alpha} \cos \theta
\end{bmatrix} \left[\begin{array}{c}
v_1^R \\
v_2^R
\end{array}\right] \\
&= \begin{bmatrix}e^{i \alpha} \cos \theta v_1^R+e^{i \beta} \sin \theta v_2^R \\
-e^{-i \beta} \sin \theta v_1^R+e^{-i \alpha} \cos \theta v_2^R\end{bmatrix} 
\end{align}
If the original eigenvector ratio was $M = \frac{v_1^R(z)}{v_2^R(z)}$, the new eigenvector ratio $M'$ becomes
\begin{equation}
    M' = \frac{e^{i \alpha} \cos \theta M+e^{i \beta} \sin \theta}{-e^{-i \beta} \sin \theta M+e^{-i \alpha} \cos \theta}.
\end{equation}
This is a special case of the Mobius transformation $M' = \frac{aM +b}{cM + d}$, where $a,b,c,d$ are complex constants. 

Mobius transformations that are rotations on the Riemann sphere can be written in the form
\begin{equation}
    M' = \frac{aM+b}{- b^* M + a^*}
\end{equation}
where $a a^* + b b^* =1$. We see that our Mobius transformation is of this form so unitary transforms of $H$ are also rotations in the $M$-Riemann sphere.

Finally, by comparing with Eq.~\eqref{Eq:unitary}, we set $\alpha = 0$, $\beta = \pi - \varphi_\Gamma$ and $\theta = \theta_\Gamma/2$. In this case the $M=\infty$ point is rotated into $M'=e^{i\varphi_\Gamma}\cot\theta_\Gamma/2$. Applying Eq.~\eqref{Eq:MRiemannSpherical} then confirms that $M'$ is the endpoint of $\mathbf{d}_\Gamma$. Similarly, the $M=0$ point is rotated into the endpoint of $-\mathbf{d}_\Gamma$.

%

\end{document}